\documentclass{article}
\usepackage{arxiv}
\usepackage[normalem]{ulem}
\usepackage{algorithm}
\usepackage{algpseudocode}
\usepackage{color} 
\definecolor{mygreen}{RGB}{28,172,0} 
\definecolor{mylilas}{RGB}{170,55,241}
\usepackage{comment}
\usepackage[utf8]{inputenc} 
\usepackage[T1]{fontenc}    
\usepackage[english]{babel}
\usepackage{hyperref}       
\usepackage{url}            
\hypersetup{
    colorlinks=true,
    linkcolor=blue,     
    citecolor=red,      
    filecolor=magenta,      
    urlcolor=red,
}
\usepackage{booktabs}       
\usepackage{amsfonts}       
\usepackage{nicefrac}       
\usepackage[table,xcdraw]{xcolor}
\usepackage{microtype}      
\usepackage{lipsum}
\usepackage{amsmath,amssymb}  
\usepackage{amsthm}         
\usepackage{graphicx}
\usepackage{caption}
\usepackage{subcaption}
\usepackage{sectsty} 
\subsubsectionfont{\normalfont\itshape} 
\usepackage{natbib}
\usepackage{fontenc}
\usepackage{multirow}
\usepackage{graphicx}

\newtheorem{theorem}{Theorem}[section]

\newtheorem{proposition}[theorem]{Proposition}

\newtheorem*{remark}{Remark}

\title{Approximations to ultimate ruin probabilities with a Wiener process perturbation}

\author{
  Yacine Koucha \\
  Brunel University London \\
  UB8 3PH Uxbridge, United Kingdom \\
  \texttt{yacine.koucha@brunel.ac.uk} \\
   \And
  Alfredo D. Egidio dos Reis  \\
  Universidade de Lisboa, ISEG \\
  1249-078 Lisboa, Portugal \\
  \texttt{alfredo@iseg.ulisboa.pt} \\
  }

\begin{document}
\maketitle


\begin{abstract}
In this paper, we adapt the classic Cramér-Lundberg collective risk theory model to a perturbed model by adding a {Wiener process} to the {compound Poisson process}, which can be used to incorporate premium income uncertainty, interest rate fluctuations and changes in the number of policyholders. {Our study is part of a Master dissertation, our aim is to make a short overview and present additionally some new approximation methods for the infinite time ruin probabilities for the perturbed risk model}. We present four different approximation methods for the perturbed risk model. The first method is based on iterative upper and lower approximations to the maximal aggregate loss distribution. The second method relies on a four-moment exponential De Vylder approximation. The third method is based on the first-order Padé approximation of the Renyi and De Vylder approximations. The last method is the second order Padé-Ramsay approximation. These are generated by fitting one, two, three or four moments of the claim amount distribution, which greatly generalizes the approximations. We test the precision of approximations using a combination of light and heavy tailed distributions for the individual claim amount. We assess the ultimate ruin probability and present numerical results for the exponential, gamma, and mixed exponential claim distributions, demonstrating the high accuracy of these four methods. Analytical and numerical methods are used to highlight the practical implications of our findings.
\end{abstract}


\section{Introduction}\label{Chapter1}

Ruin theory, section of risk theory and 
%
a field of mathematics, is an important part of actuarial education with application to non-life insurance, uses mathematical models to explain an insurer’s level on vulnerability to ruin. Risk theory, has its origins in the early 20-th century, when Filip Lundberg published his 1903 paper on the classical surplus process, \cite{lundberg_1903}. \cite{andersen_1957} adapted Lundberg’s process to allow for other claim inter-arrival times. As such, key quantities of interest are the ruin probabilities, either finite or infinite, distribution of surplus immediately prior to ruin, the deficit at the time of ruin,  dividend problems, time to ruin, {giving more common topics in the non-life actuarial literature. }

{Giving a summarised but more self-contained presentation, this is part of a master dissertation,} we present the standard model {as given by \cite{bowers_2000}, the classical Cramér-Lundberg risk model,} 
\begin{equation} \label{CramerRiskProcess}
    U(t)=u+ct-S(t), \quad S(t)=\sum^{N(t)}_{i=0} X_{i}, \quad  t\geq0\,,
\end{equation}
{where $X_0\equiv 0$, $U(t)$ is the surplus at time $t$, $u=U(0)\geq 0$ is the initial capital or reserve,
}
$c$ ($>0$) is the rate at which premiums are received, $S(t)$ is the aggregate claim amounts occurred in $(0,t]$, $N(t)$ is the number of claims up to time $t$ and $X_{i}$ is the individual claim amount $i$. We {consider} the counting process $\left\{N(t),  t\geq0\right\}$ as a Poisson process with intensity rate $\lambda>0$ and so $\left\{S(t) : t\geq0\right\}$ is a compound Poisson process. The {sequence  $\left\{X_{i}\right\}_{i=1}^\infty$ is a set of independent and identically distributed random variables}, with cumulative distribution function (CDF), $F_{X}(\cdot)$, such that $F_{X}(0)=0$ and the $k$-th ordinary moment $\mu_{k}=\mathbb{E}[X^{k}]$, which we assume to exist, for some $k\in\mathbb{N}$. The model assumes that $\left\{X_{i} \right\}_{i=1}^\infty$ and $\left\{N(t) , t\geq0\right\}$ are independent. {Also, for the model to have economic sense it is usually assumed that there exists some positive safety loading, such that 
\begin{equation}
    \theta=c(\lambda \mu_{1})^{-1}-1>0,
\end{equation}
is a strictly positive loading coefficient. 
This is known as the {\it income condition}.}
Otherwise $c<\lambda \mu_{1}$ and so this risk business would be {ultimately negative with probability one.} This is done in order to ensure that ruin does not arise with certainty. As shown by \cite{alcoforado2021ruin} many results, formulae, in ruin theory are {mathematically survive beyond this condition, however we keep it here as we deal only with ruin probabilities}.

{Perturbed processes are becoming much more relevant, since they can describe the observed reality in financial markets with greater accuracy than the classical model in Equation (\ref{CramerRiskProcess})}. {As far as non-life insurance and their modelling is concerned, there is extensive literature on the so called perturbed risk process}, with many contributions in this field, in particular recent developments such as: the Wiener process (also known as the Brownian motion), the $\alpha$-stabled process, the general diffusion risk processes, \cite{thorin_1974}, the perturbed compound Poisson risk process with investment, \cite{yin_wang_2008} and the geometric Lévy process, \cite{wang_chen_yang_gao_2018}. The biggest drawback in the original perturbed risk process, see \cite{dufresne_gerber_1991}, was that the Brownian motion was not sufficient to model big changes and differences. \cite{furrer_1998} remedied this by proposing a further generalisation to the perturbed process. After a thorough review of the literature, it seems that Lévy processes has been restricted to a Brownian motion and an $\alpha$-stable process.  The reader must be familiar with works on risk and ruin theory. As such, stochastic calculus, statistics, renewal theory and probability theory are key areas of interest. A great contemporary book to read is available in \cite{klugman_panjer_willmot_2019}. 

The aim of this manuscript is to propose and compare approximations for the probability of the process ever falling into ruin (i.e. ultimate ruin probabilities) using a mixture of light and more heavy-tailed claim distributions, for a particular risk model perturbed by a Wiener process. We particularly follow and extend the ideas by  \cite{seixas_reis_thesis} and \cite{seixas_egidio2013}. Ruin probabilities have been shown to be exponential functions when claim sizes follow an exponential distribution, see \cite{asmussen_albrecher_2010}. The idea of approximating empirical data in the form of ordinary moments is like bread and butter of classical statistics and probability. Some well-known approximations used in modern risk theory are discussed here, such as an explicit Pollaczek-Khinchine formula for the Laplace transform, as well as approximations and extensions to numerous works by \cite{dufresne_gerber_1989,vylder_marceau_1996,grandell_2000,avram_chedom_horvath_2011}, which all fit a high number of ordinary moments of the claim amount distribution. 

We present four different approximation methods for the perturbed risk model. The first method is based on iterative upper and lower approximations to the maximal aggregate loss distribution. The second method relies on a four-moment exponential De Vylder approximation. The third method is based on the one-point Padé approximation of the Renyi and De Vylder approximations. The last method is the second order Padé-Ramsay approximation. These are achieved by fitting one, two, three or four moments of the claim amount distribution, and thus generalising these approximations considerably. We use a combination of light and heavy tailed distributions for the individual claim amount to test the precision of approximations. Since input data is usually linked to uncertainty, it is interesting to develop approximations based on the finite number of ordinary moments formed by the expansion of the Laplace transform power series around zero. 


The manuscript is organised as follows. In Section~\ref{Chapter_insurance_model}, we present the perturbed risk model and derive some essential results. In Section~\ref{Chapter_Ruin_Probability}, we introduce common ruin elements that bind this work together, namely the ultimate ruin probability in infinite time, the adjustment coefficient, a decomposition of the ruin probability, the maximal aggregate loss random variable and the Pollaczek-Khinchine formula. In Section~\ref{Chapter_Approximations}, we present four different approximation methods for the perturbed risk model using the approaches outlined in the previous sections. In Section~\ref{Chapter_Results}, we use numerical approximations to test our hypothesis on the validity and accuracy of each approximation method. Finally, Section~\ref{chapter_conclusion} closes with a discussion on our results, new findings, concluding remarks, recommendations and a possible future work. Computations were carried out using the following programming softwares: \textit{Mathematica}, \textit{MATLAB} and \textit{Microsoft Excel}.
\section{The Model}\label{Chapter_insurance_model}
In this section, we summarise the perturbed model by adding  another source of randomness to model in Equation~\eqref{CramerRiskProcess}, i.e.~the Brownian motion with a drift component. We consider this section as a spiritual sequel to the work presented in \cite{seixas_reis_thesis,seixas_egidio2013}. In practice, there seems to be two approaches when it comes to applying Brownian motions, that is by (i) replacing the aggregate claim process, or (ii) using Brownian motion as perturbation to the classical model. We are interested in the second approach. 


\subsection{The perturbed risk process} \label{Sec:PerturbedRiskProcess}

Inspired by risk theory applications, we present the perturbed risk process $\{V(t):t\geq0\}$, where it is assumed that the process $U(t)$ and the Wiener process $\{W(t):t\geq0\}$ are independent, and so, the model at time $t$ is given by:
\begin{equation} \label{perturbed_process_equation}
V(t)=U(t)+\sigma W(t), \quad t\geq0 \,,
\end{equation}
where $V(t)$ is an extension to the classical model in (\ref{CramerRiskProcess}) {with the inclusion of a perturbation given by a Wiener process (see \cite{durrett_2019}) and a diffusion coefficient $\sigma$ ($>0$) which expresses an additional uncertainty for aggregate claims and premium income}. {The Wiener process is a stochastic process, defined on a complete probability space $(\Omega, \mathcal{F}, \mathbb{P})$, and characterised by the following properties:
    \begin{enumerate}
        \item $W(0)=0$;
        \item $\left\{W(t):t\geq0\right\}$ has stationary, independent increments for $0\leq t_{1}<t_{2}<\dots<\infty$;
        \item $W(t)$ is almost surely continuous in $t$;
        \item $W(t+s)-W(s)$ follows a Gaussian distribution with mean zero and variance $t$, i.e. $\mathcal{N}(0,t)$.
    \end{enumerate}
{$W(t)$ is well-defined for moments greater than zero because it is Gaussian distributed. In later sections (i.e. Section~\ref{Chapter_Approximations}) in order to match moments, we need certain of them to exist up to a certain order, for the claim amount distribution. The Wiener process is a well-known Lévy process that is regularly encountered in applied mathematics and actuarial science. In general, a Levy process $Z_{\alpha}(t)$ is a generalisations of stochastic processes with stationary, independent increments. The Wiener process is the result of the intersection of the Gaussian process class and the $\alpha$-stable Levy process class when $\alpha=2$, i.e. $Z_2(t)=W(t)$. Other known distributions includes Cauchy distribution ($\alpha=1$) and the Levy distribution ($\alpha=0.5$), see for instance  \cite{papoulis_pillai_2014}}.
\subsection{Deriving the central moments of $V(t)$}
\subsubsection{Cumulants of $W(t)$ and $S(t)$}
We start by computing the cumulants of $W(t)$ and $S(t)$ which are necessary to derive the central moments of $V(t)$. The cumulants are well characterized in actuarial literature, \cite{simar_1976,applebaum_2004}. {Assuming the existence of the cumulant generating function (CGF), a convenient way to obtain the $k$-th cumulant of a random variable is by taking the $k$-th derivative CGF of that random variable, evaluated at $s=0$. A CGF is the natural logarithm of the moment-generating function (MGF), which existence is implicit. The first three cumulants are equal to the mean ($k=1$), variance ($k=2$) and third central moment ($k=3$), respectively.} However, higher-order integer cumulants from $k\geq 4$ {do not correspond to similar central moments}, but rather more complicated polynomial functions of moments. 

Since $W(t)$ is distributed by a Gaussian $\mathcal{N}(0,t)$ with MGF $M_{W(t)}(s)=\exp(ts^2/2)$, then the CGF is given by
\begin{equation}\label{CGF_Wt}
    \varphi_{W(t)}(s)=\ln\left[\exp(ts^2/2)\right]=ts^2/2, \quad s\in\mathbb{R}.
\end{equation}
Here, the second cumulant of $W(t)$ is $t$ and the rest are zero. Since $S(t)$ follows the compound Poisson distribution as seen in (\ref{CramerRiskProcess}), the CGF is given by
\begin{equation}\label{CGF_compound_poisson}
    \varphi_{S(t)}(s)=\ln\left[{M_N(\ln{M_X(s)})}\right]=\lambda t(M_X(s)-1), \quad s\in\mathbb{R},
\end{equation}
where $M_X(s)$ is the MGF of the claims distribution. By taking the $k$-th derivative of (\ref{CGF_compound_poisson}) and setting $\mu_k=M^{(k)}_X(0)$, the $k$-th cumulants of $S(t)$ are equal to $\lambda t\mu_{k}$ for $k\in\mathbb{N}$.

\subsubsection{Central moments of $V(t)$}\label{central_moments_Vt}
{We can now calculate the central moments of $V(t)$ using (\ref{CGF_Wt}) and (\ref{CGF_compound_poisson}) which will be used in later sections. The CGF of $V(t)$ simplifies to, due to independence between $U(t)$ and $W(t)$,
    \begin{eqnarray*}
        \varphi_{V(t)}(s)
        &= &\ln \mathbb{E}[\exp{\{sV(t)\}}] 
        \\
        &=& s(u+ct)+\varphi_{S(t)}(-s)+\varphi_{W(t)}(\sigma s) \notag\\
        &=& s(u+ct)+\lambda t(M_X(-s)-1)+t(s\sigma)^2/2. \label{CGF_Vt}
    \end{eqnarray*}
    }
    Note that $M_X(-s)$ is a Laplace transform. Thus, taking the first three derivatives for $s$ and setting $s=0$ produces the respective central moments for the first three terms:
    \begin{equation}\notag
        \nu_1=u+ct-\lambda t\mu_1, \quad \nu_2=\lambda t\mu_2+\sigma^2 t, \quad \nu_3=-\lambda t\mu_3.
    \end{equation}
    Higher-order integer cumulants from $k\geq 4$ are not the same as moments about the mean. Hence, central moments $\nu_k$ will take the form
    \begin{equation} \label{Upsilon12345}
        \nu_k= \varphi^{(k)}_{V(t)}(0)+f_k(\nu_i), \quad i,k\in\mathbb{N},
    \end{equation}
    where $f_k(\nu_i)$ is a polynomial function of central moments $\nu_i$, equal to $3\nu_2^2$ and $10\nu_3 \nu_2$ for the fourth and fifth central moments, respectively, but zero otherwise for the first three central moments. Now, using the deduction produced in (\ref{Upsilon12345}), the fourth and fifth central moments will generate:
\begin{equation}\notag
    \nu_4=\lambda t\mu_4+3\sigma^4 t^2+3(\lambda t\mu_2+\sigma^2 t)^2, \quad \nu_5=-\lambda t\mu_5-10\lambda t\mu_3 (\lambda t\mu_2+\sigma^2 t).
\end{equation}

\section{Ruin Probability Methods} \label{Chapter_Ruin_Probability}

In this section, we introduce common ruin elements that bind this work together for the model presented in (\ref{perturbed_process_equation}), namely, the ultimate ruin probability in infinite time (\autoref{section_ruin_common}), an upper bound approximation to the ruin probability using the adjustment coefficient (\autoref{section_ruin_adjustment}), a decomposition of the ruin probability due to the individual claim amount and the oscillation (\autoref{section_ruin_decomp}), the maximal aggregate loss random variable (\autoref{section_ruin_maximal}) and the Pollaczek-Khinchine formula using an approximation method to calculate the ultimate ruin probability (\autoref{section_ruin_PK}). All stochastic quantities are defined on a complete probability space.

\subsection{The ultimate ruin probability}\label{section_ruin_common}

For simplicity, we consider the probability of ruin in infinite time according to (\ref{perturbed_process_equation}). Let $T_u$ be the random variable representing the time when ruin occurs, from initial surplus $u$, that is:
\begin{equation} \label{Tu}
    T_u=\inf \left\{V(t)\leq0\,:\,t\geq0\, | u\right\}, \quad u\geq0,
\end{equation}
otherwise $T_u=\infty$}, i.e.~ruin doesn't occur and $V(t)\geq 0$ $\forall$ $t\geq0$. Now, suppose we let

\begin{equation}\label{psi(u,y)}
    \Psi(u,y)=P\Big(\{ T_u<\infty\} \, \cap \, \{V(t)\in(-y,0)\} | V(0)=u\Big), \quad u,y\geq0,
\end{equation}

 be the probability that ruin occurs with initial surplus $u$ and the deficit immediately after ruin occurs is at most $y$, then setting $y\rightarrow\infty$ to (\ref{psi(u,y)}), we obtain:

\begin{equation} \label{Guy}
    \Psi(u)=P\Big(\{ T_u<\infty\} \, \cap \, \{V(t)\in(-\infty,0)\} | V(0)=u\Big)=:\mathbb{P}(T_u<\infty), \quad u\geq0,
\end{equation}

where $\Psi(u)$ is the ultimate ruin probability in continuous time and infinite time horizon with the universal boundary condition $\Psi(\infty)=0$. 
{We denote its derivative by $\psi(u)=\frac{d}{du}\Psi(u)$.}

Using equations (\ref{Tu}) and (\ref{Guy}), we define $\overline{\Psi}(u)=1-\Psi(u)$ as the survival or non-ruin probability, i.e. the probability that ruin never occurs from initial surplus $u$. {We will see later that $\overline{\Psi}(u)$ also corresponds to a CDF}. Now, to guarantee that $\overline{\Psi}(u)\neq0$ for all $u\geq0$, as said in Section \ref{Chapter1}, we must assume the net profit condition is positive, i.e.
\begin{equation} \label{ProfitCondition}
    c-\mu_{1}\lambda>0.
\end{equation}
This means that for each unit of time, the premium income exceeds the expected aggregate claim amount. If this condition fails, then $\overline{\Psi}(u)=0$ which leads to $\Psi(u)=1$ for all $u\geq0$. Condition (\ref{ProfitCondition}) brings economic sense to the classical model, and therefore it is convenient to write $c=(1+\theta)\mu_{1}\lambda$ for $\theta>0$. 


\subsection{An upper bound for $\Psi(u)$ using the adjustment coefficient}\label{section_ruin_adjustment}

The Cramér-Lundberg's adjustment coefficient, denoted as $R$~($>0)$, leads to the well-known Lundberg's inequality,  is a risk measure for a surplus process that is used to approximate the ruin probability in light-tailed claims (by an upper bound). Assuming (\ref{ProfitCondition}) is satisfied, then $r=R$ is the only positive solution to equation:
\begin{equation} \label{AdjustCoeffEq}
    \underbrace{\lambda(M_X (r)-1)}_{g_A(r)}=\underbrace{cr-(\sigma r)^2/2}_{g_B(r)}, \quad r<2\theta\mu_1/\mu_2,
\end{equation}
where $M_{X}(r)=\mathbb{E}[e^{rX}]$ is the MGF of the claim amount distribution, whose existence we assume.
Since the coefficient of $r^2$ of the quadratic function $g_B(r)$ is negative and depends on $\sigma\neq0$, the parabola therefore has a maximum point and opens downward faster as $\sigma$ increases, thus $\sigma$ and $R$ are inversely proportional. Moreover, the process $\exp({-R\{V(t)-u\}})$ is a martingale with mean one, see \cite{tzeng_schmidt_schmidli_teugels_rolski_2001}.
{It can also be noted that an upper bound (denoted as $\Psi_{\text{Lun+}}(u)$)  to the probability of ruin satisfies the following inequality (Lundberg inequality's)}
\begin{equation}
    \Psi(u)\leq \Psi_{\text{Lun+}}(u)=e^{-Ru}, \quad u\geq0, \quad R<2\theta\mu_1/\mu_2\,,
\end{equation}
and there exists some constant $k>0$ such that $\Psi(u)\sim ke^{-Ru}$ as $u\rightarrow \infty$.  

\begin{figure}[h]
    \centering
    \includegraphics[width=0.65\textwidth]{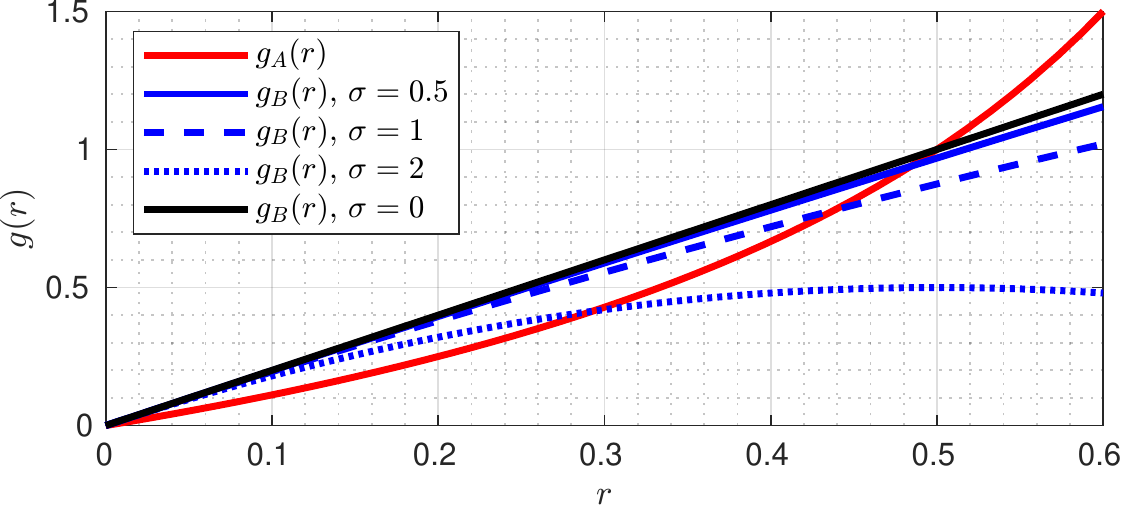}
    \caption{$R$, positive intersection between curve $g_A(r)$ and line $g_B(r)$; $\{c=2, \theta=1, \mu_1=1\}$.}
    \label{fig:Adjustment_coefficient_R}
\end{figure}

We may be interested in the first and second derivatives that offer us insight into the shape of (\ref{AdjustCoeffEq}). Let $g(r)=g_A(r)-g_B(r)=0$, on taking derivatives, we conclude that $g'(0)=\lambda \mu_1-c<0$ and $g''(0)=\lambda\mu_2+\sigma^2>0$. The first derivative implies that $g(r)$ is a decreasing function at $r=0$. However, the second derivative indicates that $r=0$ is a minimum on the range of $g(r)$ whilst concaving upwards to infinity (since $\sigma>0$ and $\lambda>0$), thus, (\ref{AdjustCoeffEq}) has a positive root at $r=R$ and a trivial solution at $r=0$. Example computations of the adjustment coefficient can be found in Figure \ref{fig:Adjustment_coefficient_R}. In this case, the claims amounts follow an exponential distribution, denoted by Exp$(1)$ with $M_X(r)=(1-r)^{-1}$, $\mu_1=1$ and $\mu_2=2$. Hence for $R<\theta$,
\begin{equation*}
    R=\frac{(\sigma^2+2c)- \sqrt{8\lambda\sigma^2+(\sigma^2-2c)^2}}{2\sigma^2}.
\end{equation*}
For instance, if $c=2$ and $\theta=1$, then $\lambda=1$. Thus, $R=(5-\sqrt{17})/2\approx0.438447<1$ when $\sigma=1$. 

\subsection{A decomposition of the ruin probability} \label{section_ruin_decomp}

Consider the perturbed process in (\ref{perturbed_process_equation}).  \cite{dufresne_gerber_1991} introduced two important decompositions of the probability of ruin: the probability of ruin when the zero line is first reached by an oscillation, denoted as $\Psi_{1}(u)$, and the probability of ruin when the zero line is first reached by a jump in an individual claim amount, denoted as $\Psi_{2}(u)$. 
Ruin due to claim is more significant than ruin due to oscillation because the shortfall at ruin in the first case can be substantial, but it is zero in the second case due to the Wiener process's continuity.

Combining these probabilities, we have the following relationship:
\begin{equation}\notag
    \Psi(u)=\Psi_1(u)+\Psi_2(u).
\end{equation}
Given that $V(0)=u$ holds, then:
\begin{equation}\notag
    \Psi_1(u)=:\mathbb{P}(T_u<\infty \; \text{and } V(t)=0), \quad \Psi_2(u)=:\mathbb{P}(T_u<\infty \; \text{and } V(t)<0).
\end{equation}
Due to the diffusive and oscillating nature of the process sample path, it follows that
\begin{equation}\notag
    \Psi_{1}(0)=\begin{cases}
        0, & u<0, \\
        1, & u=0,
    \end{cases} \quad 
    \Psi_2(0)=\begin{cases}
        1, & u<0, \\
        0, & u=0.
    \end{cases}
\end{equation}

Applying standard renewal theory techniques, \cite{dufresne_gerber_1991} arrived to a generalization that for $u\geq0$:
\begin{align}\notag
\Psi(u)=q(1-H_1(u))+(1-q)\bigg(H_1 (u)-H_3(u)\bigg)+(1-q) \int_0^u h_3(x)\Psi(u-x) \,dx,  
\end{align}
where 
\begin{equation}\notag
    h_3(z)=\int_0^z h_1(x) h_2(z-x)\,dx\equiv\int_0^z h_1(z-x) h_2(x)\,dx,
\end{equation}
is the convolution concentrated over a finite range $(0,z)$, with density functions $h_1(\cdot)$ and $h_2(\cdot)$, defined by:
\begin{equation}
h_1(x)=\tau e^{-\tau x}, \quad h_2(x)=\mu_1^{-1}\overline{F_X}(x), \quad x\geq0, \label{density_h1h2}
\end{equation}
where $\tau=2c/\sigma^2$ for $\sigma>0$. The corresponding CDFs are $H_1(\cdot)$ and $H_2(\cdot)$. It can be noted that $h_1(x)$ is an exponential PDF with mean $1/\tau$ and $h_2(x)$ is the equilibrium density function. Lastly, the two types of ruin can therefore be expressed in closed-form, i.e.
\begin{align}
\Psi_{1}(u)= \; & 1-H_1 (u)+(1-q) \int_0^u \Psi_{1} (u-x)h_3(x) \,dx, \\
\Psi_{2}(u)= \; & (1-q)(H_1 (u)-H_3(u))+(1-q) \int_0^u \Psi_2 (u-x)h_3(x) \,dx.
\end{align}
These renewal applications have provided insight into the theory, from there we can obtain numerical solutions for $\Psi(u)$, at least, see  \cite{dufresne_gerber_1991}.
\subsection{Maximal aggregate loss}\label{section_ruin_maximal}

\begin{figure}[h]
    \centering
    \includegraphics[width=\linewidth]{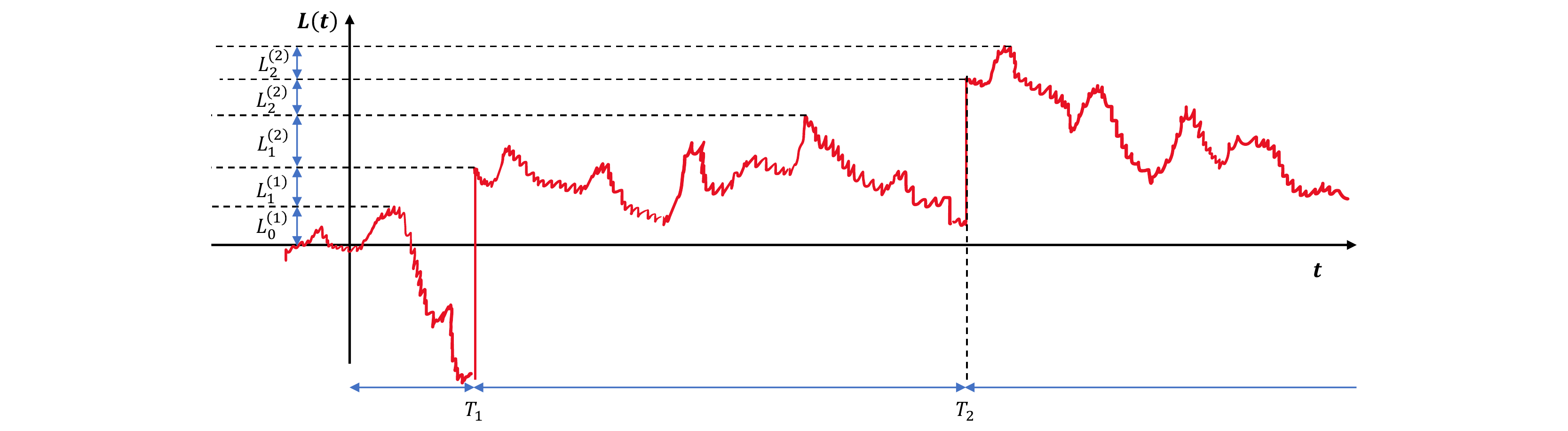}
    \caption{Decomposition of the the aggregate loss process with $\sigma>0$ and record heights $L$.}
    \label{fig:DecAggLoss}
\end{figure}

A common basis to the ruin probability approximations is the re-expression of $\Psi(u)$ in terms of the distribution of the maximal aggregate loss. We begin by defining the maximal aggregate loss variable $L=\sup{\{L(t)\,:\,t\geq0\}}$, where $L(t)=u-V(t)$, and identifying some key elements for the rest of the paper. It follows directly that
\begin{equation}\label{CDF_L}
    F_L(u)=\mathbb{P}(L(t)\leq u)=\mathbb{P}(V(t)\geq 0)=\overline{\Psi}(u), \quad t,u\geq0.
\end{equation}
The distribution of $L$ is proper and absolutely continuous if $\sigma>0$, as its PDF $f_L(0)=0$. (If $\sigma=0$ we recover the classical compound Poisson risk model and in that case the distribution is of mixed type with a probability mass at ``0''). The decomposition of $L(t)$ was first discussed in \cite{dufresne_gerber_1991}. {There is also a discussion in \cite{seixas_reis_thesis} and \cite{seixas_egidio2013}}. We now want to obtain an expression for the decomposition of $L(t)$. Working with (\ref{perturbed_process_equation}), 
the decomposition yields:
\begin{align}\label{loss_equation_complete}
    L=\max \{S(t)-c-\sigma W(t)\}=L_0^{(1)}+L_M, \quad L_M=\sum_{i=1}^{M}\bigg(L_i^{(1)}+L_i^{(2)}\bigg),
\end{align}
where 
\begin{equation*}
    L_i^{(1)}=\max\big\{L(t)\,|\,t\in(T_i,T_{i+1})\big\}-L(T_i), \quad L_i^{(2)}=L(T_i)-\Big[L(T_{i-1})+L_{i-1}^{(1)}\Big].
\end{equation*}
Here, $L_i^{(1)}$ and $L_i^{(2)}$ independent and identically distributed random variables representing the record highs due to oscillation and claim occurrences, with PDFs $h_1(\cdot)$ and $h_2(\cdot)$, respectively, and $M$ is the number of records of $L(t)$ that are due to a claim and follows a geometric distribution, with probability mass function (PMF), $\mathbb{P}(M=k)=q(1-q)^k$ for $k\in\mathbb{N}_0$. The PDFs $h_1(\cdot)$ and $h_2(\cdot)$ are given in (\ref{density_h1h2}). The CDF of $L$ in (\ref{CDF_L}) is therefore given by
\begin{equation*}
    F_L(u)=\sum_{k=0}^{\infty}q(1-q)H_1^{*(k+1)}*H_2^{*k}.
\end{equation*}
A visual decomposition of $L(t)$ from a typical sample path over time $t$ can be seen in Figure \ref{fig:DecAggLoss}.

We are also interested in extracting moments of $L$ by obtaining a general form for the MGF. We begin by making some modifications to (\ref{loss_equation_complete}). 
Let $L_i^{(3)}=:L_i^{(1)}+L_i^{(2)}$ and 
$L_M=: \sum_{i=1}^{M} L_i^{(3)} $ so that $L_M$ is a compound geometric random variable. Since $M$ and $L_i^{(1)}$ are geometrically and exponentially distributed, respectively (they are independent, also) then their respective MGF's are given by
\begin{equation}\notag
    M_M(r)=\frac{q}{1-(1-q)e^r}, \quad M_{L_i^{(1)}}(r)=\frac{\tau}{\tau-r}, 
    \quad r\in\mathbb{R},
\end{equation}
%
The MGF of $L_i^{(2)}$ can be obtained by using the expected value definition of a MGF: 
\begin{align}\label{MGF_L2}
    M_{L_i^{(2)}}(r)&=\int_0^{\infty}e^{rx}h_2(x)\,dx=\mu_1^{-1}\Bigg[\int_0^{\infty}e^{rx}\,dx-\int_0^{\infty}e^{rx}F_X(x)\Bigg]\,dx=(\mu_1 r)^{-1}[M_X(r)-1].
\end{align}
Therefore, the MGF for $L_i^{(3)}$ is
\begin{equation}\label{MGF_L3}
    M_{L_i^{(3)}}(r)=M_{L_i^{(1)}}(r)M_{L_i^{(2)}}(r)=\frac{\tau}{\mu_1(\tau-r)r}[M_X(r)-1].
\end{equation}
Since, $L_M$ is compound geometric random variable, the corresponding MGF is given by
\begin{equation}\notag
    M_{L_M}(r)=P_M(M_{L_i^{(3)}}(r))=q\Big[1- \frac{\tau(1-q)}{\mu_1(\tau-r)r}[M_X(r)-1]\Big]^{-1},
\end{equation}
where $P_M(z)$ is the probability generating function (PGF) of $M$ at point $z$. A useful relationship relating the PGF and MGF is $P_M(z)=M_M(\ln{z})$. Hence, the MGF of $L$ is
\begin{align}
    M_{L}(r)=M_{L_0^{(1)}}(r)M_{L_M}(r)
    &=\frac{q r \tau \mu_1}{r(\tau-r)\mu_1 + (q-1)\tau(M_X(r)-1)}\, .\label{MGF_L}
\end{align}
\subsection{The Pollaczek-Khinchine formula}\label{section_ruin_PK}
The Pollaczek-Khinchine formula was first published in \cite{pollaczek_1930}, where a major study on queueing theory was conducted; this formula explains the relationship between the queue length and a time distribution, by taking Laplace transforms for an M/G/1 queue (i.e. where jobs follow a Poisson process). All data about possible ruin probabilities is summarized in the Pollaczek-Khinchine formula for their Laplace transformations and is sought to calculate the ultimate ruin probability, \cite{avram_banik_horvath_2018}. In this section, we derive the Pollaczek-Khinchine formula with a new approximation method to calculate the ultimate ruin probability. 

We need to first obtain an analytical expression for the  Lévy–Khintchine/Laplace exponent of our perturbed process $V(t)$ using methods outlined by 
\cite{avram_banik_horvath_2018}. The Lévy–Khintchine/Laplace exponent of ruin process is given by
\begin{equation}\notag
    \hat{V}(s)=cs-s\lambda(1-M_X(-s))+s^2\sigma^2/2,
\end{equation}
where $M_X(-s)=f^*_X(s)$ is the Laplace transform of the claim's PDF. The variance component $s^2\sigma^2/2$ is obtained from the Laplace transform of the Brownian motion $W(t)$. Here we use a superscript (*) to denote that a Laplace transform has taken place. It may be more useful to rewrite the transformed PDF as a function of its CDF, i.e. $f^*_X(s)=sF^*_X(s)$, \cite{bracewell_2000,feller_1971}.

From 
\cite{avram_chedom_horvath_2011,avram_banik_horvath_2018}, if we replace component $r$ with $-s$ in the maximal aggregate loss expression in (\ref{loss_equation_complete}); then we can form a Laplace transform of aggregate loss random variable which coincides with the Pollaczek-Khinchine formula for the Laplace transformed ruin PDF $\psi^*(s)$ i.e.
\begin{align}\label{PK_formula}
    \psi^*(s)&=\mathbb{E}[L_0^{(1)}+L_M]=1-s\Psi^*(s), \notag\\
    \Longleftrightarrow \psi^*(s)&=\frac{s}{\hat{V}(s)}\lim_{s\rightarrow0}\Bigg[\frac{d}{ds}\hat{V}(s)\Bigg]=\frac{q}{1-(1-q)h_2^*(s)+s/\tau}.
\end{align}
Hence,
\begin{equation}\label{PK_inverse_laplace_ruin_probability}
    \Psi^*(s)=\frac{s+\tau(1-q)(1-h_2^*(s))}{s(s+\tau-\tau(1-q)h_2^*(s))}.
\end{equation}

The function $\Psi^*(s)$ emphasizes that the result in the perturbed case depends only on $h_2^*(x)$. Hence, the Laplace transform of $h_2(x)$ can be given by
\begin{equation}\label{LT_h2_MGF}
    h_2^*(s)=\int_0^{\infty}e^{-sx}\mu_1^{-1}[1-F_X(x)]\,dx
    =\mu_1^{-1}\Big[1/s-F_X^*(s)\Big]=\frac{1-M_X(-s)}{s\mu_1}.
\end{equation}
For the sake of simplicity, if we set $\sigma\rightarrow0$ and so $\tau\rightarrow\infty$, then (\ref{PK_formula}) provides a lovely rendering of the Pollaczek-Khinchine formula, which can be extended to a geometric sequence, giving rise to
\begin{equation}\notag
    \psi^*(s)=\frac{q}{1-(1-q)h_2^*(s)}=q\sum_{k=0}^{\infty}[(1-q)h_2^*(s)]^k.
\end{equation}
The {rationale} behind this is that $\psi^*(s)$ is revealed to be the Laplace transform of a geometric sum on convolutions of the equilibrium distribution. Similarly, we end up with the same distribution as the maximal aggregate loss in (\ref{loss_equation_complete}). The behavior of $\psi^*(s)$ as $s\rightarrow\infty$ differentiates between the perturbed ($\sigma>0$) and non-perturbed ($\sigma=0$) case, i.e.
    \begin{equation}\label{limiting_behaviour_psi}
        \lim_{s\rightarrow\infty}[1-\psi^*(s)]
        =\lim_{s\rightarrow\infty}{s\Psi^*(s)}=\lim_{u\rightarrow0}\Psi(u)=
        \begin{cases}
            1-q, & \text{if $\sigma=0$}, \\
            1, & \text{if $\sigma>0$}.
        \end{cases}
    \end{equation}
Since we derived the MGF for $L(t)$ in (\ref{MGF_L}), we can then go one step further and obtain factorial reduced moments, which are found by normalizing with respect to the exponential moments of $L$. We will now define a one-point Padé approximation of Laplace transforms (i.e. Renyi and De Vylder) to be used later in this paper. We redefine the Padé approximation by the given notation and application seen in 
 \cite{avram_chedom_horvath_2011,avram_banik_horvath_2018}:
\begin{equation}\label{pade_approx}
    \mathcal{P}_{(n-1,n)}(\psi^*)^{(n)}(s)=:\mathcal{P}(\psi^*|s,n), \quad n=0,1,\dots
\end{equation}
where $\mathcal{P}(\psi^*|s,n)$ is the truncated formal power series and denotes the classical Padé approximation based on the Taylor series around zero, with integer $n$. Padé approximations can be applied to divergent summation series up to $2n-1$. The purpose behind this approximation is that the Renyi and De Vylder approximations, are assumed to be the one-point Padé approximations of $\Psi^*(s)$ around the “zero-th” Taylor point, of orders $(n-1,n)$ at $n=1$. 
 
\section{Main Approximation Methods} \label{Chapter_Approximations}

In this section, we provide our main approximation methods. The first method is based on iterative upper and lower approximations to the maximal aggregate loss distribution (\autoref{section_upper_lower}). The second method relies on a four-moment exponential De Vylder approximation (\autoref{section_de_vylder_approx}). The third method is the Padé approximation of first order (\autoref{sec_pade_1}). The last method is the Padé-Ramsay approximation of second order (\autoref{sec_pade_2}).

\subsection{Upper and lower approximations to the maximal aggregate loss}\label{section_upper_lower}

We extend the work set out in \cite{dufresne_gerber_1989} which was later updated in 
\cite{seixas_reis_thesis,seixas_egidio2013}. We define new random variables from the maximal aggregate loss random variables in (\ref{loss_equation_complete}) with $L_0^{j,(1)}\equiv L^j$ if $M=0$ for bound $j=\{-,+\}$. Each $L_i^{j,(k)}$, for $k=1,2$, must be concentrated on a positive lattice $\vartheta\mathbb{N}_0=\{0,\vartheta,2\vartheta,3\vartheta\dots\}$ where the lattice width $\vartheta>0$. In this application, $L^{-,(k)}=\vartheta\big[L_i^{(k)}/\vartheta \big]$ and $L^{+,(k)}=\vartheta\big[L_i^{(k)}/\vartheta+1 \big]$ for $i=0,1,\dots,M$, reducing expression (\ref{loss_equation_complete}) to
\begin{equation}\label{loss_equation_approximation}
    L^j=L_0^{j,(1)}+\sum_{i=1}^M 
    \big(L_i^{j,(1)}+L_i^{j,(2)}\big).
\end{equation}
Each summand of $L$ approximates the lower and upper multiples of $\vartheta$, that is, $L\in(L^-,L^+)$; this leads to bounds for the ruin probability $\Psi(u)$, i.e.
\begin{equation}
    \Psi_{\text{DG}}^{-}(u)\leq\Psi(u)\leq\Psi_{\text{DG}}^{+}(u), \quad u=0,1,2,\dots,
\end{equation}
where $\Psi_{\text{DG}}^j(u)=1-\mathbb{P}(L^j\leq u)$. Let $L_i^{j,(3)}=L_i^{j,(1)}+L_i^{j,(2)}$ denote the sum of the loss random variables with probability density function $p_n^j(\cdot)$ for bound $j=\{-,+\}$. In the context of actuarial practice, the discretization of claims are useful for maximal aggregate loss random variables. Now, for a suitably small $\vartheta$, the probability of obtaining an lower and upper difference, is
\begin{align}
    p_n^-(\vartheta)=P\Big(L_i^{-,(3)}&=\vartheta n\Big)=
    \left.
        \begin{cases}
            H_3(\vartheta), & \quad n=0 \\
            H_3(\vartheta(n+1))-H_3(\vartheta n), & \quad n=1,2,\dots \\
        \end{cases}
    \right\} \\
    p_n^+(\vartheta)=P\Big(L_i^{+,(3)}&=\vartheta n\Big)=
    \left.
        \begin{cases}
            0, & \quad n=0 \\
            H_3(\vartheta n)-H_3(\vartheta(n-1)), & \quad n=1,2,\dots \\
        \end{cases}
    \right\}
\end{align}
where $H_3(\cdot)$ is the convolution CDF concentrated on positive numbers. Note that the CDF of $L_i^{j,(3)}$ is suitably arithmetic. {The probability functions (PF's)} of $L^-$ and $L^+$ can be obtained using the {\cite{panjer1981recursive}'s recursion formula
under a compound geometric distribution, see also 
\cite{klugman_panjer_willmot_2019}}. We are interested in the compound random variable $L$. Since the frequency distribution $M$ is geometrically distributed  with parameter $q$, and $L_i^{j,(3)}$ takes values on the non-negative integers, then the PF of $L^j$, denoted by $\mathbb{P}(L^j=\vartheta n)=g^j(\vartheta n)=:g_n^j$, is given by
\begin{equation}
    g_n^j=\frac{1-q}{1-(1-q)p_0^j}\sum_{i=1}^n p_i^j g_{n-i}^j, \quad n=1,2,\dots
\end{equation}
with initial probability
\begin{equation}
    g_0^j=P_M(p_0^j)=\frac{qp_0^j}{1-(1-q)p_0^j},
\end{equation}
where $P_M(p_0^j)$ is the probability generating function of $M$ at the point $p_0^j=\mathbb{P}(L_i^{j,(3)}=0)$. Since the bounded maximal aggregate loss has CDF $\mathbb{P}(L^j\leq n)=\sum_{k=0}^{n} g_k$, we can then use the Panjer recursion for $n\geq 0$ to obtain bounded compound probabilities, i.e.
\begin{equation}
    g_0^-=\frac{qp_0^-}{1-(1-q)p_0^-}, \quad g_0^+=q, \quad
    g_n^-(\vartheta)=g_0^-\sum_{i=1}^n p_i^- g_{n-i}^-, \quad g_n^+(\vartheta)=(1-q)\sum_{i=1}^n p_i^+ g_{n-i}^+.
\end{equation}
Hence, the exact ruin probability is bounded by
\begin{equation}\label{ruin_prob_bounds}
    1-\sum_{k=0}^n g_k^-(\vartheta)\leq\Psi(\vartheta)\leq 1-\sum_{k=0}^n g_k^+(\vartheta).
\end{equation}
Evaluating at these boundaries will prove to be effective in testing the precision of other approximations for situations where we do not have exact figures for the ultimate ruin probability. 


\subsection{De Vylder approximation}\label{section_de_vylder_approx}

The main idea behind De Vylder’s approximation technique was to replace the classical risk process $U(t)$ with a new process (and new parameters) by using a three-moment exponential approximation, say, $U_{\text{3ME}}(t)$ with mean $1/\beta$. {Let's} consider a new perturbed process, characterized by replacing $V(t)$ with a four-moment approximation to $V(t)$, say $V_{\text{4ME}}(t)$ with new parameters $c_*$ and $\sigma^2_*$. We also replace $S(t)$ with  $S_{\text{4ME}}(t)$ which is a new compound Poisson process with parameter $\lambda_*$ and $X(t)$ is replaced with an exponential distributed random variable $X_{\text{4ME}}(t)\sim\text{Exp}(\beta)$ with mean $1/\beta$. The raw moments of Exp($\beta$) are calculated using $\mu_{k,\text{4ME}}=\Gamma(1+k)/\beta^k$ for $k\in\mathbb{N}$, where $\Gamma(a)$ is the gamma function. The central moments of $V(t)$ were derived in Section \ref{central_moments_Vt} and are presented in Table \ref{tab:central_moments_4MEDV} with the central moments of $V_{\text{4ME}}(t)$. The key idea is to match the first four central moments using the relationship given by $\nu_k=\nu_{k,\text{4ME}}$ for $k=1,\dots,4$. For example, when $k=1$ we have $ct-\lambda t\mu_1=c_* t-\lambda_* t/\beta$.

\begin{table}[h]
\caption{The first four central moments of $V(t)$ and $V_{\text{4ME}}(t)$.}
\label{tab:central_moments_4MEDV}
\centering
\begin{tabular}{@{}rcc@{}}
\toprule
$k$ & $\nu_k$ & $\nu_{k,\text{4ME}}$ \\ \toprule
\multicolumn{1}{|c|}{1} & \multicolumn{1}{c|}{$ct-\lambda t\mu_1$} & \multicolumn{1}{c|}{$c_* t-\frac{\lambda_* t}{\beta}$} \\ \midrule
\multicolumn{1}{|c|}{2} & \multicolumn{1}{c|}{$\sigma^2 t+\lambda t\mu_2$} & \multicolumn{1}{c|}{$\sigma_*^2 t+\frac{2\lambda_* t}{\beta^2}$} \\ \midrule
\multicolumn{1}{|c|}{3} & \multicolumn{1}{c|}{$-\lambda t\mu_3$} & \multicolumn{1}{c|}{$\frac{-6\lambda_* t}{\beta^3}$} \\ \midrule
\multicolumn{1}{|c|}{4} & \multicolumn{1}{c|}{$6t\sigma^4+6t\lambda\sigma^2 \mu_2+3t\lambda^2 \mu_2^2+\lambda\mu_4$} & \multicolumn{1}{c|}{\begin{tabular}[c]{@{}c@{}}$6t\sigma_*^4+\frac{12t\lambda_* \sigma_*^2}{\beta^2}+\frac{12t\lambda_*^2+24\lambda_*}{\beta^4}$\end{tabular}} \\ \midrule
\multicolumn{1}{l}{} & \multicolumn{1}{l}{} & \multicolumn{1}{l}{}
\end{tabular}
\end{table}

Solving all four equations simultaneously using Table~\ref{tab:central_moments_4MEDV} and $\nu_k=\nu_{k,\text{4ME}}$ for $\lambda_*$, $c_*$, $\sigma_*$ and $\beta$ yields
\begin{equation}\notag
    \lambda_*=\frac{32\lambda\mu_3^4}{3\mu_4^3}, \quad c_*=\lambda\Bigg(\frac{8\mu_3^3}{3\mu_4^2}+\theta\mu_1\Bigg), \quad 
    \sigma^2_*=\lambda\Bigg(\mu_2-\frac{4\mu_3^2}{3\mu_4}\Bigg)+\sigma^2, \quad \beta=\frac{4\mu_3}{\mu_4}\,.
\end{equation}
%
\cite{dufresne_gerber_1991} devised the following method for determining the ultimate ruin probability. If the claim amount distribution is from a combination of a family of exponential distributions with PDF $f_X(x)=\sum_{k=1}^n w_k f_k(x)$ with weights $w_i>0$ such that $\sum_{i=1}^n w_i=1$ and parameters $\beta_i>0$, then the exact ruin probability is:
    \begin{equation} \label{ExactRuin4E}
        \Psi_{\text{4ME}}(u)=\sum_{k=1}^{n+1} C_k e^{-r_k u}, \quad u\geq0,
    \end{equation}
    where
    \begin{equation}\notag
        C_k=\prod_{j=1}^n (r_k-\beta_j)/\beta_j \cdot \prod_{j=1, j\neq k}^{n+1} r_j/(r_k-r_j), \quad k=1,2,\dots,n+1,
    \end{equation}
    with $\sum_{k=1}^n C_k=1$ and $r_1,r_2,\dots,r_n$ being the solutions of ${w_1}/{(\beta_1-r)}+\dots+{w_n}/{(\beta_n-r)}={(2c_*-\sigma^2_* r)}/{(2\lambda_*)}$. 
The ruin probability in (\ref{ExactRuin4E}) allows us to extract an infinite number of different distributions within the family of exponentials. In this paper, however, we shall consider a straightforward case. Suppose we set $n=1$ in (\ref{ExactRuin4E}); this ensures that our claim amount distribution is exponentially distributed with parameter $\beta$ and $w=1$, and that our probability of ruin is a mixture of two exponentials, i.e.
\begin{equation}\label{ruin_probability_4ME}
    \Psi_{\text{4ME}}(u)=C_1 e^{-r_1 u}+C_2 e^{-r_2 u},
\end{equation}
where
\begin{equation}\label{C1_C2_r}
    C_1=\frac{(r_1-\beta)r_2}{\beta(r_1-r_2)}, \quad
    C_2=\frac{(r_2-\beta)r_1}{\beta(r_2-r_1)}, \quad
    \frac{\lambda_*}{\beta-r}=c_*-\frac{\sigma_*^2 r}{2}.
\end{equation}
Solving the RHS equation in (\ref{C1_C2_r}) for $r$ leads to
\begin{equation*}
    r_{1,2}=\frac{(2c_*+\beta\sigma_*^2)\pm\sqrt{4(c_*^2-\beta c_* \sigma_*^2+2\lambda_* \sigma_*^2)+\beta^2 \sigma_*^4}}{2\sigma^2_*}, \quad \sigma_*^2>0.
\end{equation*}

\cite{grandell_2000} showed that the approach above gives the precise ruin probability for exponential or gamma claims, as well as extremely good approximations for other distributions with four moments. 
\cite{burnecki_teuerle_2011} also provided numerical illustrations to show that this method improves on De Vylder's ruin probability, which is known for being the "best" among standard approximation techniques.

\subsection{One-point Padé approximations}\label{sec_pade_1}

Cramér-Lundberg, De Vylder and Renyi's classical ruin theory approximations are all one-point Padé approximations. In perspective of advances in computing, we think that the current literature does not optimise the potential of the Padé approximations. A key point demonstrated in the current literature is that Padé approximations  do not work well with heavy-tailed claim distributions (even if we match a single moment) around a non-zero positive point, \cite{avram_banik_horvath_2018}. Below we will extend the current literature with improved approximations.

\subsubsection{A Padé approximation to the Renyi approximation}

Consider a two-moment Renyi exponential approximation, which is from the family of Ramsay-type approximations of $h_2(x)$, in \cite{avram_chedom_horvath_2011,avram_banik_horvath_2018}. Since we can consider this as a Padé approximation of the aggregate loss PDF at $n=1$, as defined in (\ref{pade_approx}), which also satisfies the limiting behavior from (\ref{limiting_behaviour_psi}), then we can obtain an approximation for the ruin probability using the Laplace transform
\begin{equation}\notag
    \Psi^*(s)\approx\frac{1-q}{s+R_0}\rightarrow h_2^*(s)\approx\frac{R_0/q}{s+R_0/q}, \quad R_0>0,
\end{equation}
where $R_0=q/\Tilde{\mu}_k$ is the Renyi coefficient and $\Tilde{\mu}_k$ is the $k$-th factorial moment, given by
\begin{equation}\notag
    \Tilde{\mu}_k=\frac{\mu_{k+1}}{\mu_1(k+1)}, \quad k=1,2,\dots,
\end{equation} 
For instance, the first factorial moment equal to $\Tilde{\mu}_1=\mu_2/(2\mu_1)=q/R_0$, and so, our two-moment Renyi exponential approximation is
\begin{equation}
    \Psi_{\text{Ren}}(u)\approx(1-q)e^{-R_0 u}, \quad R_0={2q\mu_1}/{\mu_2}>0,
\end{equation}
which satisfies the constraint $\Psi_{\text{Ren}}(0)=1-q$. Note that the Renyi coefficient is bounded by the adjustment coefficient, $0<R_0<R=2\theta\mu_1/\mu_2$ since $q<\theta$, and this bound tightens when $q\rightarrow\theta$. Hence, Renyi's approximation to the ruin probability is guaranteed to be
\begin{equation}\notag
    \Psi_{\text{Ren}}(u)\leq\Psi_{\text{Lun+}}(u), \quad u\geq0.
\end{equation}
This can also be regarded as a simplified version of the Beekman-Bowers' approximation, \cite{grandell_2000}, which leads us to believe that this method is probably not as good as De Vylder’s exponential approximation in (\ref{ruin_probability_4ME}) since there we matched four moments and here we only matched two.

\subsubsection{A Padé approximation to the De Vylder approximation}

In \autoref{section_de_vylder_approx}, De Vylder's approximation was used to match the first four moments of the perturbed risk process to the exponential claims distribution. However, we will now use the factorial moments of the aggregate loss density to demonstrate that this estimate matches the expansion sequence provided by the Padé approximation. We start by expanding the Pollaczek-Khinchine formula deduced in (\ref{PK_inverse_laplace_ruin_probability}) in power series:
\begin{equation}\label{de_vylder_PK_power_series}
    \Psi^*(s)=\frac{\frac{\eta_{2,\sigma}}{2!}-s\frac{\eta_{3}}{3!}+s^2\frac{\eta_4}{4!}-\dots}{\rho+s\frac{\eta_{2,\sigma}}{2!}-s^2\frac{\eta_{3}}{3!}+s^3\frac{\eta_4}{4!}-\dots}\approx\sum_{j=1}^n\frac{A_j}{s+\beta_j}, \quad j=1,2,\dots.
\end{equation} 
The parameters $\eta_k=\lambda\mu_k$ for $k=1,2,\dots$ represent moments of a Lévy measure (with $\eta_{2,\sigma}=\lambda\mu_2+\sigma^2$), $A_j$ are constants, $\beta_j$ are exponential parameters and $\rho=cq>0$ is the profit parameter. For simplicity, we will consider a one-point De Vylder approximation, which reduces the RHS of (\ref{de_vylder_PK_power_series}) to $A_1/(s+\beta_1)=:A_*/(s+\beta_*)$. Hence, taking the inverse Laplace transform of (\ref{de_vylder_PK_power_series}) yields the desired ruin probability:
    \begin{equation}\label{ruin_probability_PKDV4}
        \Psi_{\text{PKDV4}}(u)\approx A_* \exp(-\beta_* u),
    \end{equation}
        where 
    \begin{equation}\notag
        A_*=\frac{2 (\sigma^2+\lambda \mu_2) \mu_3}{2 (\sigma^2+\lambda \mu_2) \mu_3+c q \mu_4}, \quad \beta_*=\frac{4 c q \mu_3}{2 (\sigma^2+\lambda \mu_2) \mu_3+c q \mu_4}.
\end{equation}
\begin{proof}
    Solving for $A_1$ and $\beta_1$ consists of obtaining two equations. Therefore, by manipulating (\ref{de_vylder_PK_power_series}) into the following
    \begin{equation}\notag
        A_1\Big(\rho+s\frac{\eta_{2,\sigma}}{2!}-s^2\frac{\eta_{3}}{3!}+s^3\frac{\eta_4}{4!}-\dots\Big)\approx
        (s+\beta_1)\Big(\frac{\eta_{2,\sigma}}{2!}-s\frac{\eta_{3}}{3!}+s^2\frac{\eta_4}{4!}-\dots\Big),
    \end{equation}
    We can then match coefficients for any power of $s$ to obtain multiple equations. In this case, we will match the zeroth and second powers of $s$ to obtain solutions with claim amount moments up to $\mu_4$:
    \begin{equation}\label{pair_A1_beta1}
        O(\text{const.}):\quad 2 A_1 c q=\beta_1 (\sigma^2+\lambda \mu_2) , \quad O(s^2): \quad 4(A_1-1) \lambda \mu_3+\beta_1 \lambda \mu_4=0 .
    \end{equation}
    Hence solving the pair of equations in (\ref{pair_A1_beta1}) simultaneously for $A_1$ and $\beta_1$ completes the proof.
\end{proof}
\begin{remark}
    Setting $\eta_{2,\sigma}=\eta_{2}$, where $\sigma=0$, reduces the above results to
    \begin{equation}\notag
        A_*=\frac{\lambda (3\mu_2\mu_4-4 \mu_3^2)}{6 (2 \lambda \mu_2 \mu_3+c q \mu_4)}, \quad \beta_*=\frac{6 \lambda \mu_2^2+4 c q \mu_3}{2 \lambda \mu_2 \mu_3+c q \mu_4}.
    \end{equation}
\end{remark}
\begin{remark}
    Under similar assumptions, we can obtain numerous approximations to the ruin probability. For instance, if we matched coefficients up to $s$ only, we would get an approximation which includes moment up to $\mu_3$ only, i.e.
     \begin{equation}\label{ruin_probability_PKDV3}
        \Psi_{\text{PKDV3}}(u)=A_3 e^{-\beta_3 u}, \quad A_3=\frac{3(\sigma^2+\lambda\mu_2)^2}{3(\sigma^2+\lambda\mu_2)^2+2cq\lambda\mu_3}, \quad \beta_3=\frac{6cq(\sigma^2+\lambda\mu_2)}{3(\sigma^2+\lambda\mu_2)^2)+2cq\lambda\mu_3}.
    \end{equation}
    However, matching coefficients up to $s^3$ will enable us to get a ruin probability approximation which includes the fifth raw moment $\mu_5$, i.e.
    \begin{equation}\label{ruin_probability_PKDV5}
        \Psi_{\text{PKDV5}}(u)=A_5 e^{-\beta_5 u}, \quad A_5=\frac{5(\sigma^2+\lambda\mu_2)\mu_4}{5(\sigma^2+\lambda\mu_2)\mu_4+2cq\mu_5}, \quad \beta_5=\frac{10cq\mu_4}{5(\sigma^2+\lambda\mu_2)\mu_4+2cq\mu_5}.
    \end{equation}
\end{remark}

\begin{table}[h]
\caption{Approximations to De Vylder's ruin probabilities, comparing models (\ref{ruin_probability_4ME}), (\ref{ruin_probability_PKDV4}), (\ref{ruin_probability_PKDV3}) and (\ref{ruin_probability_PKDV5}). Claim amounts are Exp(1) distributed with $\sigma=0.5$.}
\label{tab:PKDV345_errors_sigma_half}
\centering
\resizebox{\textwidth}{!}{%
\begin{tabular}{@{}ccccccccccc@{}}
\toprule
$u$ & 0.0 & 1.0 & 2.0 & 5.0 & 10.0 & 20.0 & 35.0 & 50.0 & 75.0 & 100.0 \\ \midrule
\multicolumn{1}{|c|}{$\Psi_{\text{4ME}}(u)$} & \multicolumn{1}{c|}{1.000000} & \multicolumn{1}{c|}{0.983435} & \multicolumn{1}{c|}{0.974799} & \multicolumn{1}{c|}{0.949347} & \multicolumn{1}{c|}{0.908394} & \multicolumn{1}{c|}{0.831713} & \multicolumn{1}{c|}{0.728655} & \multicolumn{1}{c|}{0.638367} & \multicolumn{1}{c|}{0.512056} & \multicolumn{1}{c|}{0.410738} \\ \midrule
\multicolumn{1}{|c|}{$\Psi_{\text{PKDV3}}(u)$} & \multicolumn{1}{c|}{0.992161} & \multicolumn{1}{c|}{0.983449} & \multicolumn{1}{c|}{0.974814} & \multicolumn{1}{c|}{0.949361} & \multicolumn{1}{c|}{0.908407} & \multicolumn{1}{c|}{0.831724} & \multicolumn{1}{c|}{0.728664} & \multicolumn{1}{c|}{0.638374} & \multicolumn{1}{c|}{0.512061} & \multicolumn{1}{c|}{0.410742} \\ \midrule
\multicolumn{1}{|c|}{$\Psi_{\text{PKDV4}}(u)$} & \multicolumn{1}{c|}{0.991189} & \multicolumn{1}{c|}{0.982495} & \multicolumn{1}{c|}{0.973877} & \multicolumn{1}{c|}{0.948473} & \multicolumn{1}{c|}{0.907597} & \multicolumn{1}{c|}{0.831054} & \multicolumn{1}{c|}{0.728171} & \multicolumn{1}{c|}{0.638025} & \multicolumn{1}{c|}{0.511892} & \multicolumn{1}{c|}{0.410694} \\ \midrule
\multicolumn{1}{|c|}{$\Psi_{\text{PKDV5}}(u)$} & \multicolumn{1}{c|}{0.992063} & \multicolumn{1}{c|}{0.984221} & \multicolumn{1}{c|}{0.976441} & \multicolumn{1}{c|}{0.953467} & \multicolumn{1}{c|}{0.916372} & \multicolumn{1}{c|}{0.846455} & \multicolumn{1}{c|}{0.751454} & \multicolumn{1}{c|}{0.667115} & \multicolumn{1}{c|}{0.547055} & \multicolumn{1}{c|}{0.448602} \\ \midrule
\end{tabular}%
}
\end{table}

\begin{table}[h]
\caption{Approximations to De Vylder's ruin probabilities, comparing models (\ref{ruin_probability_4ME}), (\ref{ruin_probability_PKDV4}), (\ref{ruin_probability_PKDV3}) and (\ref{ruin_probability_PKDV5}). Claim amounts are Exp(1) distributed with $\sigma=1$.}
\label{tab:PKDV345_errors_sigma_1}
\centering
\resizebox{\textwidth}{!}{%
\begin{tabular}{@{}ccccccccccc@{}}
\toprule
$u$ & 0.0 & 1.0 & 2.0 & 5.0 & 10.0 & 20.0 & 35.0 & 50.0 & 75.0 & 100.0 \\ \midrule
\multicolumn{1}{|c|}{$\Psi_{\text{4ME}}(u)$} & \multicolumn{1}{c|}{1.000000} & \multicolumn{1}{c|}{0.989188} & \multicolumn{1}{c|}{0.982439} & \multicolumn{1}{c|}{0.963060} & \multicolumn{1}{c|}{0.931625} & \multicolumn{1}{c|}{0.871799} & \multicolumn{1}{c|}{0.789186} & \multicolumn{1}{c|}{0.714402} & \multicolumn{1}{c|}{0.605175} & \multicolumn{1}{c|}{0.512649} \\ \midrule
\multicolumn{1}{|c|}{$\Psi_{\text{PKDV3}}(u)$} & \multicolumn{1}{c|}{0.995575} & \multicolumn{1}{c|}{0.988989} & \multicolumn{1}{c|}{0.982447} & \multicolumn{1}{c|}{0.963078} & \multicolumn{1}{c|}{0.931642} & \multicolumn{1}{c|}{0.871815} & \multicolumn{1}{c|}{0.789200} & \multicolumn{1}{c|}{0.714414} & \multicolumn{1}{c|}{0.605184} & \multicolumn{1}{c|}{0.512655} \\ \midrule
\multicolumn{1}{|c|}{$\Psi_{\text{PKDV4}}(u)$} & \multicolumn{1}{c|}{0.993377} & \multicolumn{1}{c|}{0.986821} & \multicolumn{1}{c|}{0.980307} & \multicolumn{1}{c|}{0.961023} & \multicolumn{1}{c|}{0.929722} & \multicolumn{1}{c|}{0.870145} & \multicolumn{1}{c|}{0.787862} & \multicolumn{1}{c|}{0.713359} & \multicolumn{1}{c|}{0.604512} & \multicolumn{1}{c|}{0.512274} \\ \midrule
\multicolumn{1}{|c|}{$\Psi_{\text{PKDV5}}(u)$} & \multicolumn{1}{c|}{0.993377} & \multicolumn{1}{c|}{0.986821} & \multicolumn{1}{c|}{0.980307} & \multicolumn{1}{c|}{0.961023} & \multicolumn{1}{c|}{0.929722} & \multicolumn{1}{c|}{0.870145} & \multicolumn{1}{c|}{0.787862} & \multicolumn{1}{c|}{0.713359} & \multicolumn{1}{c|}{0.604512} & \multicolumn{1}{c|}{0.512274} \\ \midrule
\end{tabular}%
}
\end{table}

\begin{table}[h]
\caption{Approximations to De Vylder's ruin probabilities, comparing models (\ref{ruin_probability_4ME}), (\ref{ruin_probability_PKDV4}), (\ref{ruin_probability_PKDV3}) and (\ref{ruin_probability_PKDV5}). Claim amounts are Exp(1) distributed with $\sigma=2$.}
\label{tab:PKDV345_errors_sigma_2}
\centering
\resizebox{\textwidth}{!}{%
\begin{tabular}{@{}ccccccccccc@{}}
\toprule
$u$ & 0.0 & 1.0 & 2.0 & 5.0 & 10.0 & 20.0 & 35.0 & 50.0 & 75.0 & 100.0 \\ \midrule
\multicolumn{1}{|c|}{$\Psi_{\text{4ME}}(u)$} & \multicolumn{1}{c|}{1.000000} & \multicolumn{1}{c|}{0.995813} & \multicolumn{1}{c|}{0.992311} & \multicolumn{1}{c|}{0.982394} & \multicolumn{1}{c|}{0.966174} & \multicolumn{1}{c|}{0.934533} & \multicolumn{1}{c|}{0.889005} & \multicolumn{1}{c|}{0.845695} & \multicolumn{1}{c|}{0.778149} & \multicolumn{1}{c|}{0.715998} \\ \midrule
\multicolumn{1}{|c|}{$\Psi_{\text{PKDV3}}(u)$} & \multicolumn{1}{c|}{0.998890} & \multicolumn{1}{c|}{0.995570} & \multicolumn{1}{c|}{0.992260} & \multicolumn{1}{c|}{0.982398} & \multicolumn{1}{c|}{0.966178} & \multicolumn{1}{c|}{0.934538} & \multicolumn{1}{c|}{0.889009} & \multicolumn{1}{c|}{0.845699} & \multicolumn{1}{c|}{0.778152} & \multicolumn{1}{c|}{0.716001} \\ \midrule
\multicolumn{1}{|c|}{$\Psi_{\text{PKDV4}}(u)$} & \multicolumn{1}{c|}{0.996678} & \multicolumn{1}{c|}{0.993372} & \multicolumn{1}{c|}{0.990077} & \multicolumn{1}{c|}{0.980258} & \multicolumn{1}{c|}{0.964110} & \multicolumn{1}{c|}{0.932606} & \multicolumn{1}{c|}{0.887269} & \multicolumn{1}{c|}{0.844137} & \multicolumn{1}{c|}{0.776858} & \multicolumn{1}{c|}{0.714942} \\ \midrule
\multicolumn{1}{|c|}{$\Psi_{\text{PKDV5}}(u)$} & \multicolumn{1}{c|}{0.995025} & \multicolumn{1}{c|}{0.990087} & \multicolumn{1}{c|}{0.985173} & \multicolumn{1}{c|}{0.970578} & \multicolumn{1}{c|}{0.946732} & \multicolumn{1}{c|}{0.900784} & \multicolumn{1}{c|}{0.836008} & \multicolumn{1}{c|}{0.775891} & \multicolumn{1}{c|}{0.685147} & \multicolumn{1}{c|}{0.605016} \\ \midrule
\end{tabular}%
}
\end{table}

\subsection{Two-point Padé-Ramsay approximation}\label{sec_pade_2}

In order to derive the ruin probability for a two-point Padé-Ramsay approximation, we need to substitute our series expansion of $h_2^*(s)$ up to $s^2$ from (\ref{LT_h2_MGF}) into (\ref{PK_inverse_laplace_ruin_probability}). This then corresponds to the approximation of the form:
    \begin{equation}
        s\Psi^*(s)\approx\frac{s+\tau(1-q)(\frac{s\mu_2}{2\mu_1}-\frac{s^2\mu_3}{6\mu_1})}{s+\tau-\tau(1-q)(1-\frac{s\mu_2}{2\mu_1}+\frac{s^2\mu_3}{6\mu_1})}\geq\frac{s+\tau(1-q)^2(\frac{s\mu_2}{2\mu_1}-\frac{s^2\mu_3}{6\mu_1})}{s+\tau-\tau(1-q)(1-\frac{s\mu_2}{2\mu_1}+\frac{s^2\mu_3}{6\mu_1})}.\label{2PP_approximation}
    \end{equation}
    The RHS of (\ref{2PP_approximation}) certainly satisfies the limiting behavior $\lim_{s\rightarrow\infty}s\Psi^*(s)=1-q$ (which can be shown by dividing the numerator and denominator by $s^2$ and using L'Hôpital's rule then setting $s\rightarrow\infty$). We can solve the following problem:
    \begin{equation}\label{inverse_LT_linear_quad_fraction}
        \Psi^*(s)\approx\frac{1+\tau(1-q)^2(\frac{\mu_2}{2\mu_1}-\frac{s\mu_3}{6\mu_1})}{s+\tau-\tau(1-q)(1-\frac{s\mu_2}{2\mu_1}+\frac{s^2\mu_3}{6\mu_1})}\equiv\frac{a_1s+a_0}{b_2s^2+b_1s+b_0},
    \end{equation}
    where
    \begin{equation}\notag
        a_0=1+\frac{(1-q)^2\tau\mu_2}{2\mu_1}, \quad a_1=\frac{-(1-q)^2\tau\mu_3}{6\mu_1}, \quad b_0=q\tau, \quad b_1=1+\frac{a_0-1}{1-q}, \quad b_2=\frac{a_1}{1-q}.
    \end{equation}
    Finally, the inverse Laplace transformation of the RHS linear-quadratic fractional expression in (\ref{inverse_LT_linear_quad_fraction}) will yield:
    \begin{equation}
        \Psi_{\text{2PP}}(u)\approx(k_1\cosh{(\zeta u)}+k_2\sinh{(\zeta u)})e^{-\eta u}
    \end{equation}
    where 
    \begin{equation*}
        k_1=\frac{a_1}{b_2}, \quad k_2=\frac{-a_1 b_1+2 a_0 b_2}{b_2 \sqrt{b_1^2-4 b_0 b_2}}, \quad \zeta=\frac{\sqrt{b_1^2-4 b_0 b_2}}{2 b_2}, \quad
        \eta=\frac{b_1}{2b_2}.
    \end{equation*}    

\section{Numerical Results} \label{Chapter_Results} 

\begin{figure}
    \centering
    \begin{subfigure}[b]{0.325\textwidth}
         \centering
         \includegraphics[width=\textwidth]{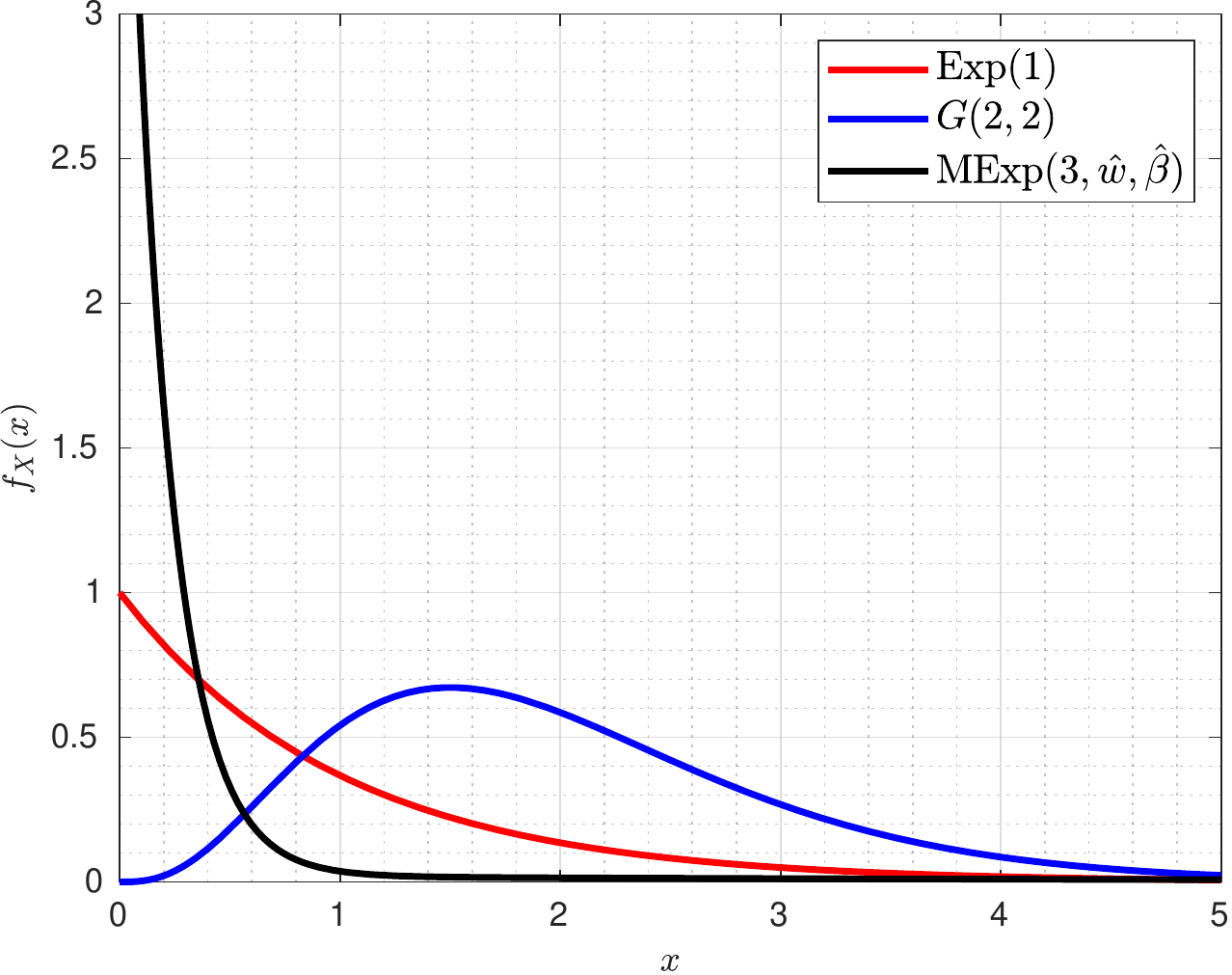}
         \caption{Density $f_X(x)$}
         \label{fig:Results_PDFs}
     \end{subfigure}
     \hfill
     \begin{subfigure}[b]{0.325\textwidth}
         \centering
         \includegraphics[width=\textwidth]{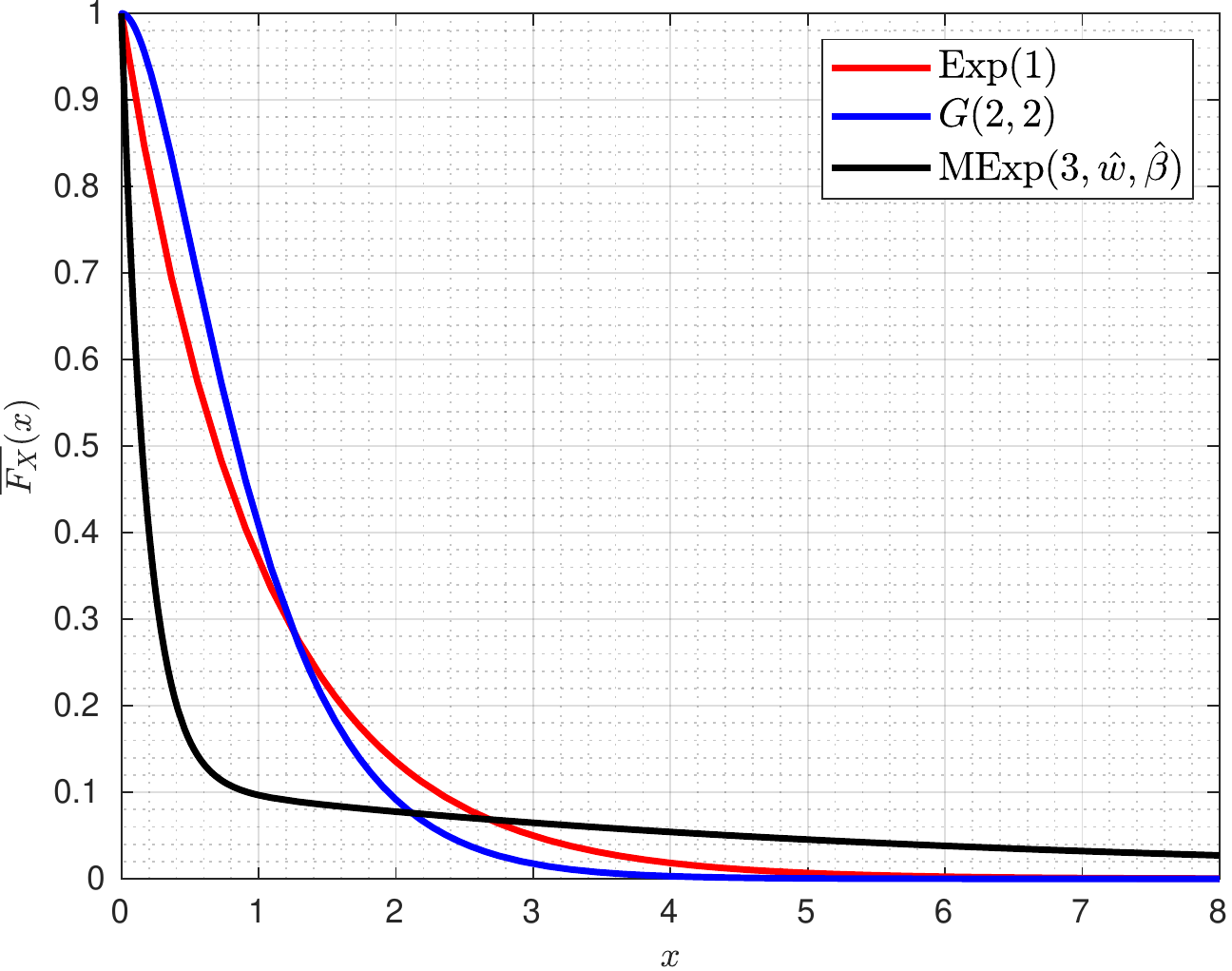}
         \caption{Tail $\overline{F_X}(x)$}
         \label{fig:Results_Tail}
     \end{subfigure}
     \hfill
     \begin{subfigure}[b]{0.325\textwidth}
         \centering
         \includegraphics[width=\textwidth]{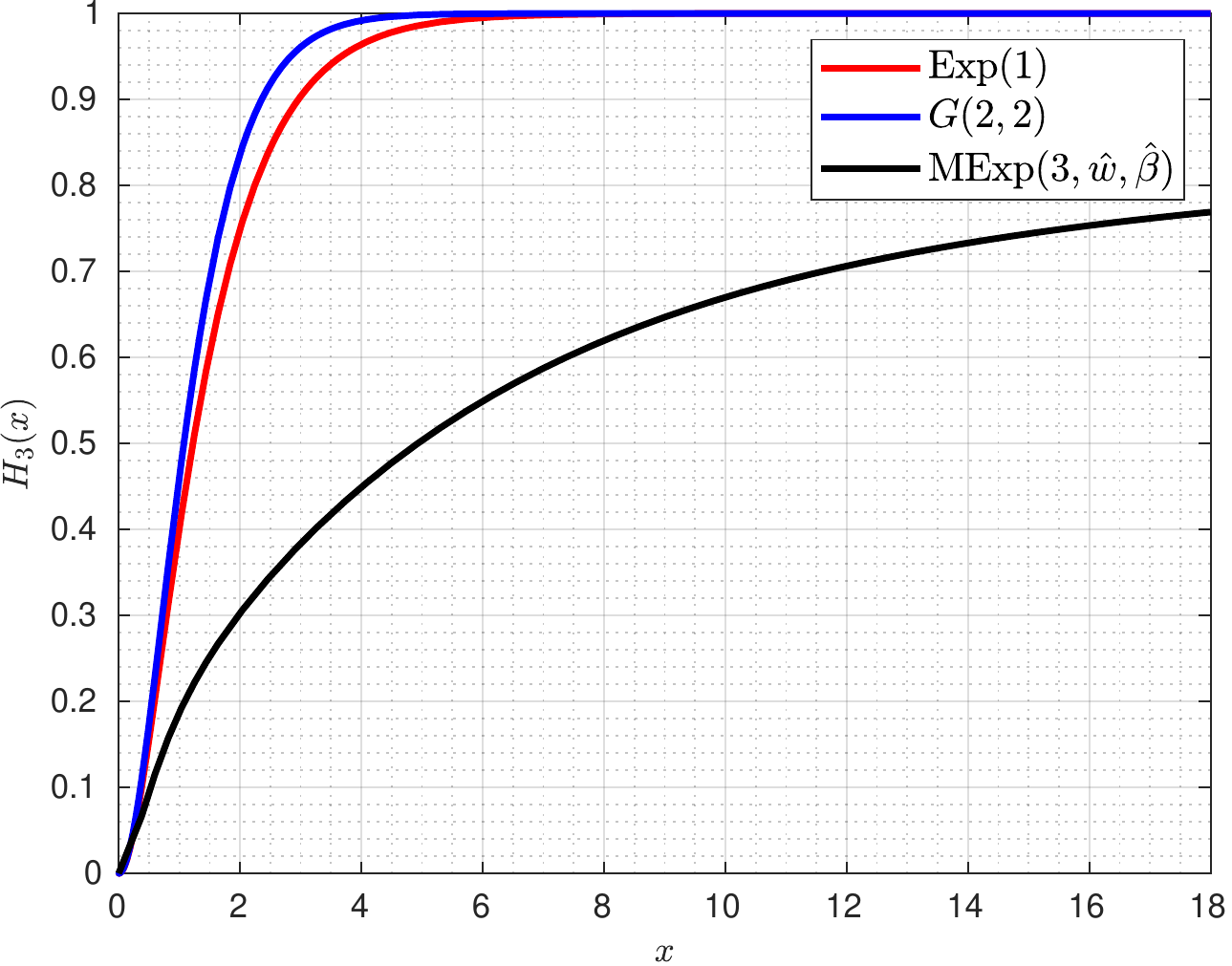}
         \caption{Convolution $H_3(x)$}
         \label{fig:Results_H3}
     \end{subfigure}
     \hfill
     \caption{Density, tail and convolution functions illustrated for Exp$(1)$, $G(2,2)$ and $\text{MExp}(3,\hat{\boldsymbol{w}},\boldsymbol{\hat{\beta}})$.}
     \label{fig:Results_PDFs_Tail_H3}
\end{figure}

In this section, we present numerous illustrations for the exact ruin probability $\Psi(u)$ and compared them to the four approximation methods $\Psi_{\text{Approx}}(u)$, namely, 
\begin{itemize}
    \item Dufresne and Gerber's bounds (i.e upper bound $\Psi_{\text{DG+}}(u)$ and lower bound $\Psi_{\text{DG-}}(u)$),
    \item De Vylder's 4-moment exponential approximation $\Psi_{\text{4ME}}(u)$,
    \item One-point Padé approximations using an inverse Laplace transformed 2-moment Renyi approximation $\Psi_{\text{Ren2}}(u)$ and a 3- and 4-moment De Vylder approximation, given by $\Psi_{\text{PKDV3}}(u)$ and $\Psi_{\text{PKDV4}}(u)$, respectively,
    \item Two-point Padé-Ramsay approximation $\Psi_{\text{2PP}}(u)$.
\end{itemize}
We assume that each claim amount distribution, denoted by $X$, has mean $\mu_1=1$, and we set $X$ to follow a gamma [model (\ref{gamma_pdf_tail})], an exponential [model (\ref{exp_pdf_tail})] and a mixture of three exponentials [model (\ref{mixed_exp_pdf_tail})]. We also set $\alpha=2$, $\lambda=1$ and $\theta=1\%$ to our perturbed process which leads to $\tau=2.02/\sigma^2$ and $q=1/101$. The exact ruin probability is calculated numerically by taking the inverse Laplace transform of (\ref{PK_inverse_laplace_ruin_probability}). Dufresne and Gerber's bounds will be limited by a fixed lattice width of $\vartheta=0.1$. Different values of $\sigma$ are observed and have been considered to observe the consistency among each ruin probability. Moreover, our equilibrium density is now equal to the claim amount's tail function, i.e. $h_2(x)=\overline{F_X}(x)$, and the convolution CDF $H_3(x)$ is updated to: 
\begin{equation*}
    H_3(x;\tau)=\int_0^x \Big(1-\exp{\{-\tau(x-t)\}}\Big)\overline{F_X}(t)\,dt, \quad x,\tau\geq0.
\end{equation*} 
$H_3(x)$ must satisfy the requirements of a distribution function
for each claim amount. 
Lastly, relative errors are used to determine the precision and accuracy of these approximations. We shall compute them using:
\begin{equation}
    \varepsilon_{\text{Approx}}(u)=|1-\Psi_{\text{Approx}}(u)/\Psi(u)|.
\end{equation} 
Figure \ref{fig:Results_PDFs_Tail_H3} presents a side-by-side comparison of the density, tail and convolution functions for each distribution of claims. Additional results for the adjustment coefficient in other distributions are available in Table \ref{tab:adjustment_coefficient_results} which will be used in determining an Lundberg upper bound for the ruin probability. We discuss our findings in the following subsections. 

\begin{table}[h]
\caption{The MGF and adjustment coefficient expressions using (\ref{AdjustCoeffEq}). Parameter settings for the mixed exponential distribution are predefined using (\ref{mixed_exp_settings}). $R$ is given for the case of $\sigma=1$.}
\label{tab:adjustment_coefficient_results}
\centering
\begin{tabular}{@{}cccc@{}}
\toprule
  & Exponential & Gamma & Mixed Exponential \\ \midrule
\multicolumn{1}{|c|}{Parameters} & \multicolumn{1}{c|}{$\beta=1$} & 
\multicolumn{1}{c|}{$k=2,\beta=2$} & 
\multicolumn{1}{c|}{$n=3,\boldsymbol{w}=\hat{\boldsymbol{w}}, \boldsymbol{\beta}=\boldsymbol{\hat{\beta}}$} \\ 
\midrule
\multicolumn{1}{|c|}{$M_X(r)$} & \multicolumn{1}{c|}{$(1-r)^{-1}$} & \multicolumn{1}{c|}{$(1-r/2)^{-2}$} & 
\multicolumn{1}{c|}{$\sum_{i=1}^3 \hat{w_i}\hat{\beta_i}/(\hat{\beta_i}-r)$} \\ 
\midrule
\multicolumn{1}{|c|}{$R$} & \multicolumn{1}{c|}{$0.0066371$} & \multicolumn{1}{c|}{$0.0079744$} & \multicolumn{1}{c|}{$0.0685513$} \\ \midrule
\end{tabular}
\end{table} 

\subsection{Gamma($k,\beta$) claims}\label{section_gamma_claims}

The two-parameter gamma distribution, denoted as $G(k,\beta)$ with shape $k>0$ and rate $\beta>0$, can be viewed as a generalization of the exponential distribution. The density $f_X(\cdot)$ and tail $\overline{F_X}(\cdot)$ function of this distribution is given by
\begin{equation}\label{gamma_pdf_tail}
    f_X(x;k,\beta)=\frac{\beta^k}{\Gamma(k)}x^{k-1}e^{-\beta x}, \quad \overline{F_X}(x;k,\beta)=1-Q(k,\beta x): \quad Q(k,\beta x)=\frac{\gamma(k,\beta x)}{\Gamma(k)}, \quad x\geq0,
\end{equation}
where $Q(k,\beta x)$ is the regularized gamma function consisting of  the gamma function $\Gamma(k)$ and the lower incomplete gamma function $\gamma(k,\beta x)$. Both functions are given by:
\begin{equation}
    \Gamma(k)=\int_0^{\infty}t^{k-1}e^{-t}\,dt, \quad \gamma(k,\beta x)=\int_0^{\beta x} t^{k-1}e^{-t}\,dt.
\end{equation}
It can be noted that $Q(k,\beta x)$ is implemented in \textit{Mathematica} as \texttt{GammaRegularized[k,0,bx]}. The distribution of $H_3(x;k,\beta,\tau)$ can only be computed numerically as it does not have a simple closed form for non-integer parameters. To ensure a mean of one, we let $k=\beta$ (we will stick with $\beta$). Therefore, the MGF of this distribution is $M_X(r)=(1-r/\beta)^{-\beta}$ with the variance of the claims determined by $1/\beta$. Raw moments can be computed easily using $\mu_j=\frac{\Gamma(j+\beta)}{\Gamma(\beta)}\beta^{-j}$ for $j+\beta>0$. 

If $k=1$, then $G(1,\beta)\equiv\text{Exp}(\beta)$ with density and tail function given by
\begin{equation}\label{exp_pdf_tail}
    f_X(x;\beta)=\beta\exp{(-\beta x)}, \quad \overline{F_X}(x;\beta)=\exp{(-\beta x)}, \quad x\geq0,
\end{equation}
Exponential distributed claims is by far the easiest to deal with in ruin theory. The hazard rate\footnote{The hazard rate (or the failure rate) is measured using $h_X(x)=f_X(x)/\overline{F_X}(x)$.} is a constant ($\beta$) with respect to time which alludes to the "memory-less" property of this distribution. Raw moments can be computed from $\mu_j=j!/{\beta^j}$ and that the variance of this distribution can be explained from $1/\beta^2$. 

\begin{figure}
    \centering
    \begin{subfigure}[b]{0.325\textwidth}
         \centering
         \includegraphics[width=\textwidth]{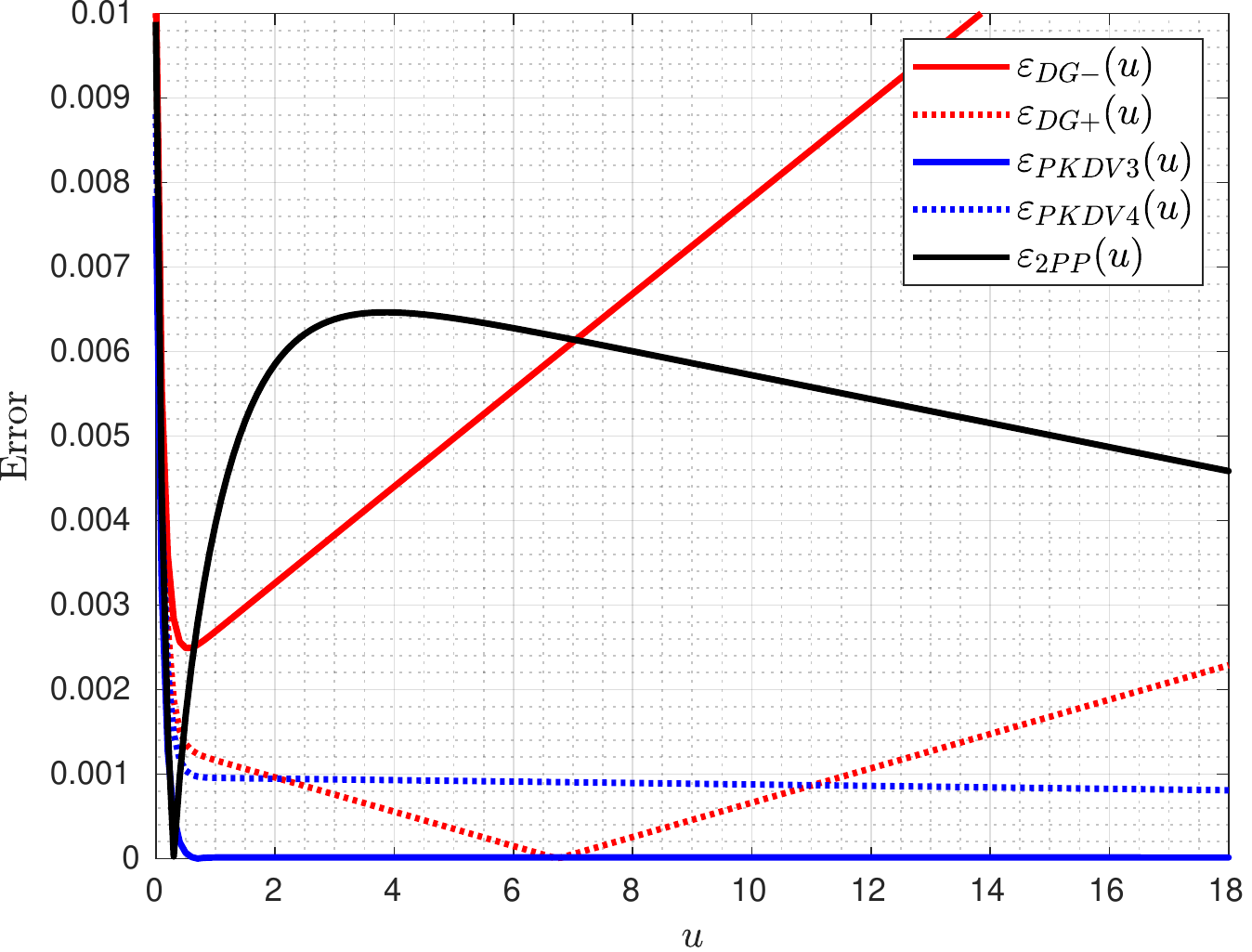}
         \caption{Exp($1$); $\sigma=0.5$}
         \label{fig:Relative_errors_exp(1)_gamma(2_2)_sigma_half_V1}
     \end{subfigure}
     \hfill
     \begin{subfigure}[b]{0.325\textwidth}
         \centering
         \includegraphics[width=\textwidth]{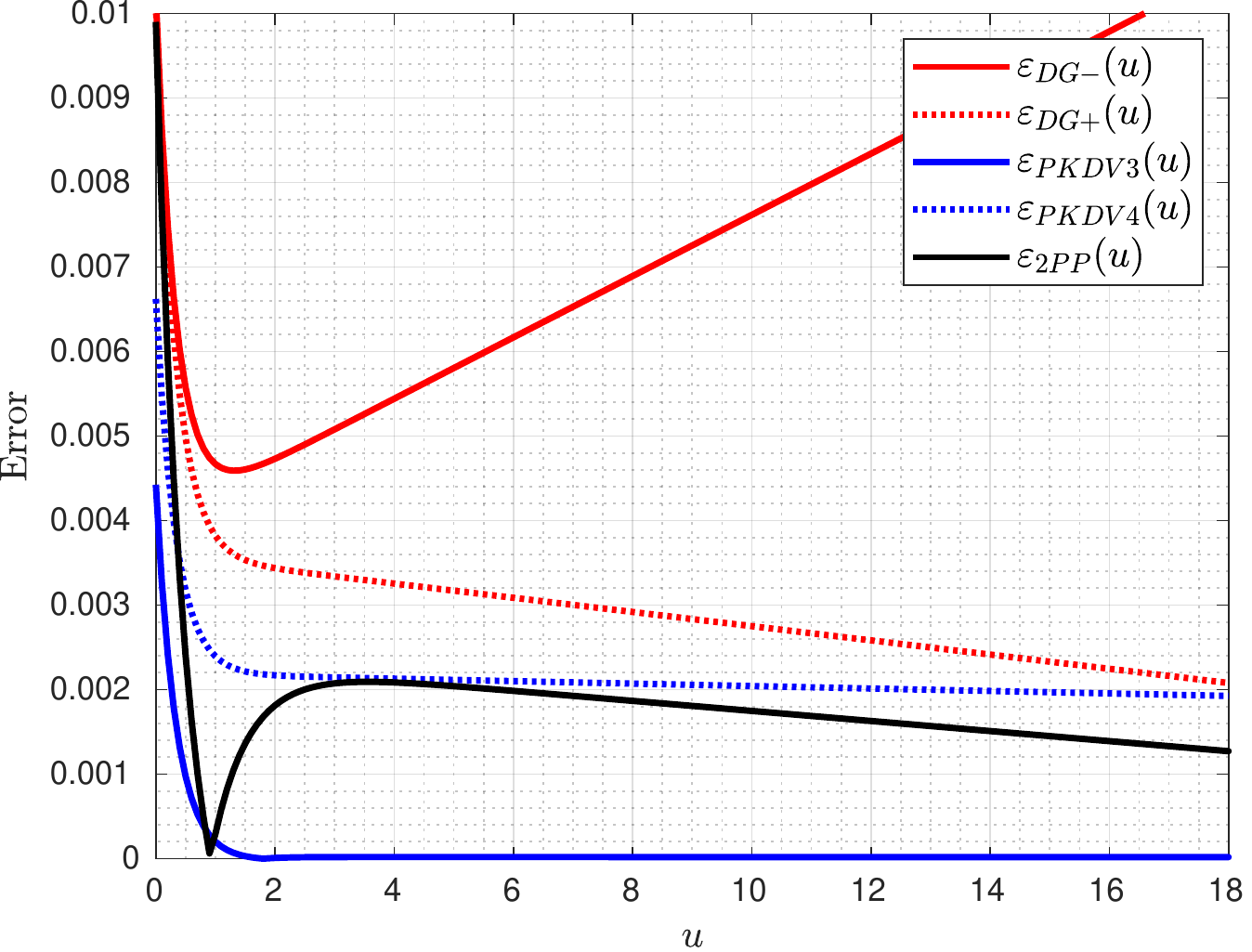}
         \caption{Exp($1$); $\sigma=1.0$.}
         \label{fig:Relative_errors_exp(1)_gamma(2_2)_sigma1_V1}
     \end{subfigure}
     \hfill
     \begin{subfigure}[b]{0.325\textwidth}
         \centering
         \includegraphics[width=\textwidth]{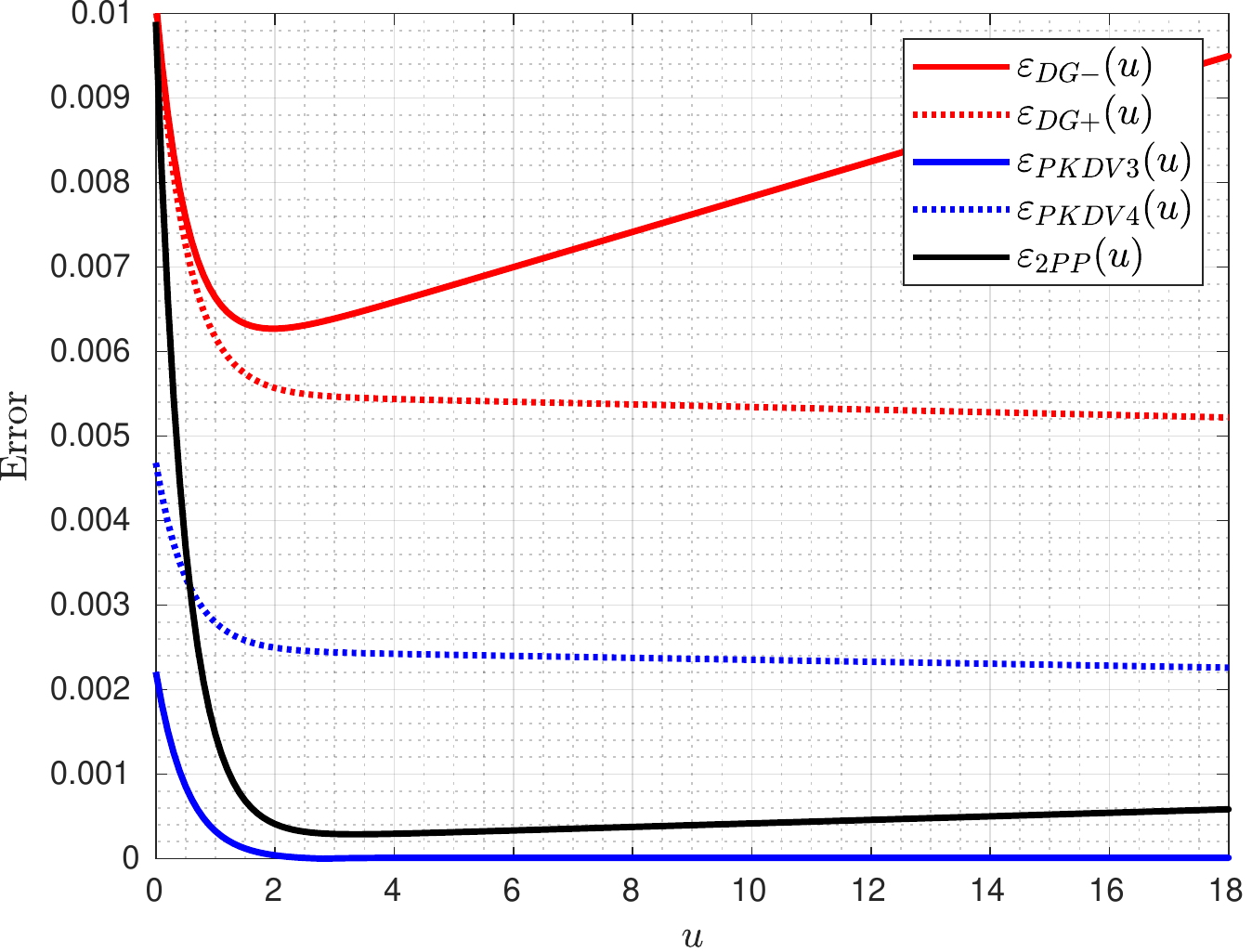}
         \caption{Exp($1$); $\sigma=1.5$.}
         \label{fig:Relative_errors_exp(1)_gamma(2_2)_sigma_onehalf_V1}
     \end{subfigure}
     \hfill
     \caption{Relative errors for approximate ruin probabilities compared to the exact ruin probability for $u\in[0,18]$ and various levels of $\sigma$. $X\sim\text{Exp}(1)$ with $\lambda=1$ and $\theta=0.01$.}
     \label{fig:Relative_errors_exp(1)_gamma(2_2)}
\end{figure}

Example calculations can be seen in Tables \ref{tab:results_exp_claims}-\ref{tab:results_gamma_claims} and Figure \ref{fig:Relative_errors_exp(1)_gamma(2_2)} which presents cases for Exp$(1)$ and $G(2,2)$ with the expected claim amount set to one. The distribution of $H_3(\cdot)$ for each case are given by:
\begin{align}
    H_{3|\text{Exp}(1)}(x;\tau)&=\frac{1-\exp{\{-\tau x\}}-\tau(1-\exp{\{-x\}})}{1-\tau}, \quad x\geq0, \\
    H_{3|G(2,2)}(x;\tau)&=1+\frac{(\tau-4)\exp{\{-\tau x\}}-\tau(\tau-3+(\tau-2)x)\exp{\{-2 x\}}}{(\tau-2)^2}, \quad x\geq0,
\end{align}        
where $H_{3|\text{Exp}(1)}(0)=H_{3|G(2,2)}(0)=0$ and $H_{3|\text{Exp}(1)}(\infty)=H_{3|G(2,2)}(\infty)=1$, satisfying the properties of a distribution function.

\begin{table}[h]
\centering
\caption{Exact and approximate ruin probabilities for $u\in(0,50]$ when $X\sim\text{Exp}(1)$ with $\sigma=1$.}
\label{tab:results_exp_claims}
\resizebox{\textwidth}{!}{%
\begin{tabular}{@{}|
>{\columncolor[HTML]{DDEBF7}}c |c|c|c|c|c|c|c|c|c|@{}}
\toprule
\textbf{$u$} &
  \cellcolor[HTML]{DDEBF7}\textbf{$\Psi(u)$} &
  \cellcolor[HTML]{DDEBF7}\textbf{$\Psi_{\text{DG–}}(u)$} &
  \cellcolor[HTML]{DDEBF7}\textbf{$\Psi_{\text{DG+}}(u)$} &
  \cellcolor[HTML]{DDEBF7}\textbf{$\Psi_{\text{4ME}}(u)$} &
  \cellcolor[HTML]{DDEBF7}\textbf{$\Psi_{\text{Ren2}}(u)$} &
  \cellcolor[HTML]{DDEBF7}\textbf{$\Psi_{\text{PKDV3}}(u)$} &
  \cellcolor[HTML]{DDEBF7}\textbf{$\Psi_{\text{PKDV4}}(u)$} &
  \cellcolor[HTML]{DDEBF7}\textbf{$\Psi_{\text{2PP}}(u)$} &
  \cellcolor[HTML]{DDEBF7}\textbf{$\Psi_{\text{Lun+}}(u)$} \\ \midrule
\textbf{0.1}  & 0.998183 & 0.989566 & 0.990099 & 0.998183 & 0.989119 & 0.994915 & 0.992720 & 0.990512 & 0.999337 \\ \midrule
\textbf{0.2}  & 0.996668 & 0.989208 & 0.989568 & 0.996668 & 0.988140 & 0.994255 & 0.992063 & 0.990775 & 0.998673 \\ \midrule
\textbf{0.5}  & 0.993242 & 0.987707 & 0.988300 & 0.993242 & 0.985210 & 0.992277 & 0.990094 & 0.990861 & 0.996687 \\ \midrule
\textbf{1.0}  & 0.989188 & 0.984570 & 0.985410 & 0.989188 & 0.980344 & 0.988989 & 0.986821 & 0.989479 & 0.993385 \\ \midrule
\textbf{1.5}  & 0.985742 & 0.981202 & 0.982259 & 0.985742 & 0.975503 & 0.985713 & 0.983558 & 0.987086 & 0.990094 \\ \midrule
\textbf{2.0}  & 0.982439 & 0.977790 & 0.979061 & 0.982439 & 0.970686 & 0.982447 & 0.980307 & 0.984220 & 0.986814 \\ \midrule
\textbf{3.0}  & 0.975929 & 0.970972 & 0.972670 & 0.975929 & 0.961123 & 0.975948 & 0.973836 & 0.977954 & 0.980286 \\ \midrule
\textbf{5.0}  & 0.963060 & 0.957469 & 0.960006 & 0.963060 & 0.942278 & 0.963078 & 0.961023 & 0.965026 & 0.967359 \\ \midrule
\textbf{10.0} & 0.931625 & 0.924528 & 0.929062 & 0.931625 & 0.896766 & 0.931642 & 0.929722 & 0.933253 & 0.935784 \\ \midrule
\textbf{25.0} & 0.843343 & 0.832351 & 0.842086 & 0.843343 & 0.773001 & 0.843358 & 0.841805 & 0.844064 & 0.847108 \\ \midrule
\textbf{50.0} & 0.714402 & 0.698694 & 0.714842 & 0.714402 & 0.603506 & 0.714414 & 0.713359 & 0.713951 & 0.717591 \\ \bottomrule
\end{tabular}%
}
\end{table}

\begin{table}[h]
\centering
\caption{Exact and approximate ruin probabilities for $u\in(0,50]$ when $X\sim G(2,2)$ with $\sigma=1$.}
\label{tab:results_gamma_claims}
\resizebox{\textwidth}{!}{%
\begin{tabular}{@{}|
>{\columncolor[HTML]{DDEBF7}}c |c|c|c|c|c|c|c|c|c|@{}}
\toprule
\textbf{$u$} &
  \cellcolor[HTML]{DDEBF7}\textbf{$\Psi(u)$} &
  \cellcolor[HTML]{DDEBF7}\textbf{$\Psi_{\text{DG–}}(u)$} &
  \cellcolor[HTML]{DDEBF7}\textbf{$\Psi_{\text{DG+}}(u)$} &
  \cellcolor[HTML]{DDEBF7}\textbf{$\Psi_{\text{4ME}}(u)$} &
  \cellcolor[HTML]{DDEBF7}\textbf{$\Psi_{\text{Ren2}}(u)$} &
  \cellcolor[HTML]{DDEBF7}\textbf{$\Psi_{\text{PKDV3}}(u)$} &
  \cellcolor[HTML]{DDEBF7}\textbf{$\Psi_{\text{PKDV4}}(u)$} &
  \cellcolor[HTML]{DDEBF7}\textbf{$\Psi_{\text{2PP}}(u)$} &
  \cellcolor[HTML]{DDEBF7}\textbf{$\Psi_{\text{Lun+}}(u)$} \\ \midrule
\textbf{0.1}  & 0.998183 & 0.989546 & 0.989806 & 0.998054 & 0.988630 & 0.996016 & 0.994233 & 0.990884 & 0.999203 \\ \midrule
\textbf{0.2}  & 0.996666 & 0.989152 & 0.989549 & 0.996523 & 0.987162 & 0.995222 & 0.993442 & 0.991320 & 0.998406 \\ \midrule
\textbf{0.5}  & 0.993199 & 0.987423 & 0.988118 & 0.993177 & 0.982774 & 0.992844 & 0.991072 & 0.991223 & 0.996021 \\ \midrule
\textbf{1.0}  & 0.988866 & 0.983660 & 0.984696 & 0.988919 & 0.975503 & 0.988893 & 0.987136 & 0.988718 & 0.992057 \\ \midrule
\textbf{1.5}  & 0.984922 & 0.979575 & 0.980918 & 0.984950 & 0.968286 & 0.984958 & 0.983215 & 0.985179 & 0.988110 \\ \midrule
\textbf{2.0}  & 0.981018 & 0.975442 & 0.977093 & 0.981027 & 0.961123 & 0.981038 & 0.979309 & 0.981355 & 0.984178 \\ \midrule
\textbf{3.0}  & 0.973235 & 0.967213 & 0.969477 & 0.973235 & 0.946954 & 0.973246 & 0.971545 & 0.973555 & 0.976361 \\ \midrule
\textbf{5.0}  & 0.957836 & 0.950962 & 0.954428 & 0.957836 & 0.919240 & 0.957847 & 0.956200 & 0.958057 & 0.960912 \\ \midrule
\textbf{10.0} & 0.920397 & 0.911522 & 0.917819 & 0.920396 & 0.853453 & 0.920407 & 0.918889 & 0.920372 & 0.923352 \\ \midrule
\textbf{25.0} & 0.816632 & 0.802749 & 0.816204 & 0.816632 & 0.683016 & 0.816640 & 0.815468 & 0.815981 & 0.819254 \\ \midrule
\textbf{50.0} & 0.669029 & 0.649540 & 0.671225 & 0.669029 & 0.471176 & 0.669035 & 0.668314 & 0.667639 & 0.671177 \\ \bottomrule
\end{tabular}%
}
\end{table}



\subsection{Mixed exponential claim distribution}\label{section_mixed_exp_claims}
In this section, we provide a crude attempt to explain the highly skewed distribution of a 1948-1951 Swedish non-industry fire insurance \cite{arfwedson_1955}, which utilizes a claim distribution for a mixture of exponentials, denoted by MExp$(n,\boldsymbol{w},\boldsymbol{\beta})$ with $n$ finite set of distribution functions, weights $w_i\geq0$ such that $\sum_{i=1}^{n}w_i=1$ and exponential parameters $\beta_i>0$. The density and tail function of this mixture distribution is given by 
\begin{equation}\label{mixed_exp_pdf_tail}
    f_X(x;n,\boldsymbol{w},\boldsymbol{\beta})=\sum_{i=1}^n w_i\beta_i e^{-\beta_ix}, \quad \overline{F_X}(x;n,\boldsymbol{w},\boldsymbol{\beta})=\sum_{i=1}^n w_i e^{-\beta_ix}, \quad x\geq0.
\end{equation} 
The parameters are predefined with the following settings (data retrieved from \cite{arfwedson_1955}):
\begin{equation}\label{mixed_exp_settings}
    n=3, \quad \boldsymbol{\hat{w}} =(0.8881815,0.1078392,0.0039793), \quad    \boldsymbol{\hat{\beta}} =(5.514588,0.190206,0.014631),
\end{equation}
which guarantees a mean of $\sum_i w_i/\beta_i\approx1$. Moreover, the MGF of this distribution is $M_X(r)=\sum_i w_i\beta_i/(\beta_i-r)$ with moments $\mu_j=j!\sum_i w_i/\beta_i^j$ for $j\in\mathbb{N}$. The variance and skewness of this distribution are very large at 42.198 and 27.687, respectively. The distribution of $H_3(\cdot)$ for this case is specified by
\begin{equation}
    H_{3|\text{MExp}(3,\hat{\boldsymbol{w}},\boldsymbol{\hat{\beta}})}(x;\tau) =
    \sum_{i=1}^3 
    \frac{\hat{w_i}(\hat{\beta_i}(1-\exp{\{-\tau x\}})-\tau(1-\exp{\{-\hat{\beta_i} x\}}))}{\hat{\beta_i}(\hat{\beta_i}-\tau)}, \quad x\geq0.
\end{equation}

\begin{table}[h]
\centering
\caption{Exact and approximate ruin probabilities for $u\in(0,50]$ when $X\sim\text{MExp}(3,\hat{\boldsymbol{w}},\boldsymbol{\hat{\beta}})$ with $\sigma=1$.}
\label{tab:results_mixed_exp_claims}
\resizebox{\textwidth}{!}{%
\begin{tabular}{@{}|
>{\columncolor[HTML]{DDEBF7}}c |c|c|c|c|c|c|c|c|c|@{}}
\toprule
\textbf{$u$} &
  \cellcolor[HTML]{DDEBF7}\textbf{$\Psi(u)$} &
  \cellcolor[HTML]{DDEBF7}\textbf{$\Psi_{\text{DG–}}(u)$} &
  \cellcolor[HTML]{DDEBF7}\textbf{$\Psi_{\text{DG+}}(u)$} &
  \cellcolor[HTML]{DDEBF7}\textbf{$\Psi_{\text{4ME}}(u)$} &
  \cellcolor[HTML]{DDEBF7}\textbf{$\Psi_{\text{Ren2}}(u)$} &
  \cellcolor[HTML]{DDEBF7}\textbf{$\Psi_{\text{PKDV3}}(u)$} &
  \cellcolor[HTML]{DDEBF7}\textbf{$\Psi_{\text{PKDV4}}(u)$} &
  \cellcolor[HTML]{DDEBF7}\textbf{$\Psi_{\text{2PP}}(u)$} &
  \cellcolor[HTML]{DDEBF7}\textbf{$\Psi_{\text{Lun+}}(u)$} \\ \midrule
\textbf{0.1}  & 0.998184 & 0.989636 & 0.989819 & 0.999675 & 0.990083 & 0.974296 & 0.970285 & 0.990099 & 0.999956 \\ \midrule
\textbf{0.2}  & 0.996675 & 0.989407 & 0.989638 & 0.999354 & 0.990066 & 0.974253 & 0.970242 & 0.990100 & 0.999912 \\ \midrule
\textbf{0.5}  & 0.993397 & 0.988680 & 0.988934 & 0.998408 & 0.990017 & 0.974124 & 0.970114 & 0.990100 & 0.999780 \\ \midrule
\textbf{1.0}  & 0.990290 & 0.987643 & 0.987872 & 0.996889 & 0.989934 & 0.973910 & 0.969901 & 0.990099 & 0.999559 \\ \midrule
\textbf{1.5}  & 0.988567 & 0.986793 & 0.987008 & 0.995439 & 0.989852 & 0.973695 & 0.969689 & 0.990096 & 0.999339 \\ \midrule
\textbf{2.0}  & 0.987440 & 0.986039 & 0.986249 & 0.994055 & 0.989770 & 0.973480 & 0.969476 & 0.990091 & 0.999118 \\ \midrule
\textbf{3.0}  & 0.985831 & 0.984661 & 0.984874 & 0.991472 & 0.989605 & 0.973051 & 0.969050 & 0.990076 & 0.998678 \\ \midrule
\textbf{5.0}  & 0.983261 & 0.982145 & 0.982373 & 0.986952 & 0.989276 & 0.972194 & 0.968200 & 0.990024 & 0.997798 \\ \midrule
\textbf{10.0} & 0.977847 & 0.976739 & 0.976994 & 0.978504 & 0.988454 & 0.970053 & 0.966076 & 0.989771 & 0.995600 \\ \midrule
\textbf{25.0} & 0.966315 & 0.965170 & 0.965447 & 0.965130 & 0.985992 & 0.963659 & 0.959734 & 0.988083 & 0.989037 \\ \midrule
\textbf{50.0} & 0.953409 & 0.952106 & 0.952394 & 0.953003 & 0.981902 & 0.953095 & 0.949257 & 0.982869 & 0.978194 \\ \bottomrule
\end{tabular}%
}
\end{table}

\subsection{Discussion}

In Tables \ref{tab:results_exp_claims}-\ref{tab:results_mixed_exp_claims}, we set $\lambda=1$, $\sigma=1$ and $\theta=0.01$ to compare the approximate ruin probabilities to their exact counterpart. All distributions had their means are equal to $1$. The rate of ruin under these distribution is proportional to its power or time, i.e. the ruin rate is constant over time. For the exponential and gamma cases, the distributions are light-tailed, and thus, the contribution of claims to ruin is expected to be less important than heavier-tailed distributions (such as the Mixed exponential case). We find that all ruin probability approximations appear to be excellent for low levels of $u$. However, only $\Psi_{\text{Ren2}}(u)$ does not fall within the limits set by $\Psi_{\text{DG}\pm}(u)$ for $u\geq1$. In general, we find that $\Psi_{\text{PKDV3}}(u)$ and $\Psi_{\text{PKDV4}}(u)$ are very consistent with De Vylder's classical four-moment exponential approximation $\Psi_{\text{4ME}}(u)$ for increasing levels of $u$. In this case, however, the 3-moment approximation is better than the 4-moment approximation of the transformed Pollaczek-Khinchine De Vylder approximation. It can be noted that $\Psi_{\text{4ME}}(u)$ is equivalent to the exact ruin probability since the claim distribution here is an exponential. 

In addition, Figures \ref{fig:Relative_errors_exp(1)_gamma(2_2)_sigma_half_V1} to \ref{fig:Relative_errors_exp(1)_gamma(2_2)_sigma_onehalf_V1} shows distinctions in the probability of ruin for each approximation under different concentrations of $\sigma$ when claims amounts follow the distribution of Exp$(1)$. Changing value of $\sigma$ has a different effect to the relative error for each approximation. For instance, the error of $\varepsilon_{\text{2PP}}(u)$ is very large for small values of $\sigma$ and increases rapidly for low levels of $u$, but then the error decreases at a linear rate when the parabola changes direction at some point on $u$; whereas for greater concentrations of $\sigma$, the error of $\varepsilon_{\text{2PP}}(u)$ is low for initial levels of $u$, but then the error rises at a slow and fairly linear pace. On the other hand, the error of $\varepsilon_{\text{PKDV4}}(u)$ works vice versa, that is, the error of $\varepsilon_{\text{PKDV4}}(u)$ increases as $\sigma$ increases; but for all levels of $\sigma$, the error decreases at a very slow and linear rate as $u$ increases. {Obviously, the contribution of the diffusion process fades away as $\sigma$ tends to zero.} Overall, most approximations to the ruin probabilities appear to be fairly close to the exact.

\section{Concluding remarks}\label{chapter_conclusion}

{In this paper, we adapted the perturbed model to the classical risk process by adding a Wiener process to the Poisson compound process that enables us to consider uncertainty about the premium revenue, interest rate fluctuations, changes in the amount of policyholders, without neglecting any other assumptions.} The findings acquired seem to give us an indication that the parameter of diffusion can have a significant impact in calculating the probability of ruin, particularly for light-tailed distributions, i.e. exponential. The illustrations for different $\sigma$ is not covered here, but one can find that the error is smaller for larger values of $\sigma$ (at least for the mixed exponential claim distribution) and for most values of $u$, a far better approximation to the ruin probability \cite{avram_chedom_horvath_2011,avram_banik_horvath_2018}.

The four approximation methods were shown to be highly accurate in the exponential, gamma, and mixed exponential cases. In particular, results have shown that the relative errors for De Vylder’s classical four moment exponential approximation, Pollaczek-Khinchine’s one-point (De Vylder case) and two-point Padé all appear capable of producing excellent results if the claim distribution is well parameterized. Dufresne and Gerber’s upper and lower bounds returned good approximations when the claim distribution was exponential. The Renyi approximation, on the other hand, produced the worst fit due to the smallest number of moments, regardless of the claim distribution chosen. It can be noted that the Renyi approximation is a simplified version of the Beekman-Bowers' approximation, see \cite{grandell_2000}.

In summary, we proposed and numerically compared efficient methods for evaluating the probability of ruin in the compound Poisson risk process perturbed by a Wiener process. We will now briefly discuss the benefits and drawbacks of each approximation method:
\begin{itemize}
    \item The upper and lower bounds method can be applied to heavy-tailed distributions as well as light-tailed individual claim amounts. It also has the benefit of putting a limit on the probability of ruin. However, it takes longer to compute than the other approximation methods.
    \item The 4-moment exponential approximation of De Vylder is certainly the quickest to compute. For light and heavy tailed distributions, it is extremely precise in computing probabilities of ruin, and if the claim amount is exponentially distributed, it is also exactly equal to the probability of ruin.
    \item The Padé approximations, like the De Vylder approximation, are computationally simple because the ruin probability is given in closed form. It's a strong method that works for both light-tailed and heavy-tailed individual claim amounts.
\end{itemize}


\section*{Acknowledgements}\label{Acknowledgements}
{Second author was partially supported by the Project CEMAPRE/REM - UIDB/05069/2020 - financed by FCT/MCTES through national funds.
}
\bibliographystyle{myapalike}
\bibliography{references}
\end{document}